\documentclass[journal=jacsat,manuscript=article]{achemso}

\usepackage{graphicx}
\usepackage{dcolumn}
\usepackage{bm}
\usepackage{hyperref}
\usepackage[mathlines]{lineno}
\usepackage{amsmath}
\usepackage{caption}
\usepackage{subcaption}
\usepackage{color}
\usepackage{import}
\usepackage{xcolor}
\usepackage{soul}

\title{Ultra-long charge carrier recombination time in methylammonium lead halide perovskites}

\author{Andr\'{a}s~Bojtor}
\affiliation{{Laboratory of Physics of Complex Matter, \'{E}cole Polytechnique F\'{e}d\'{e}rale de Lausanne, CH-1015 Lausanne, Switzerland}}
\alsoaffiliation{{Department of Physics, Institute of Physics, Budapest University of Technology and Economics, Műegyetem rkp. 3., H-1111 Budapest, Hungary}}
\alsoaffiliation{{Semilab Co. Ltd., Prielle Kornélia u. 4/a, 1117 Budapest, Hungary}}

\author{S\'{a}ndor~Kollarics}
\affiliation{{Laboratory of Physics of Complex Matter, \'{E}cole Polytechnique F\'{e}d\'{e}rale de Lausanne, CH-1015 Lausanne, Switzerland}}
\alsoaffiliation{{Department of Physics, Institute of Physics, Budapest University of Technology and Economics, Műegyetem rkp. 3., H-1111 Budapest, Hungary}}

\author{Bence~G\'{a}bor~M\'{a}rkus}
\affiliation{{Stavropoulos Center for Complex Quantum Matter, Department of Physics and Astronomy, University of Notre Dame, Notre Dame, Indiana 46556, USA}}
\alsoaffiliation{{Institute for Solid State Physics and Optics, Wigner Research Centre for Physics, Budapest H-1525, Hungary}}
\alsoaffiliation{{Department of Physics, Institute of Physics, Budapest University of Technology and Economics, Műegyetem rkp. 3., H-1111 Budapest, Hungary}}

\author{Andrzej~Sienkiewicz}
\affiliation{{Laboratory of Physics of Complex Matter, \'{E}cole Polytechnique F\'{e}d\'{e}rale de Lausanne, CH-1015 Lausanne, Switzerland}}
\alsoaffiliation{{ADSresonances SARL, Route de Gen\`{e}ve 60B, CH-1028, Pr\`{e}verenges, Switzerland}}

\author{M\'{a}rton~Koll\'{a}r}
\affiliation{{Laboratory of Physics of Complex Matter, \'{E}cole Polytechnique F\'{e}d\'{e}rale de Lausanne, CH-1015 Lausanne, Switzerland}}

\author{L\'{a}szl\'{o}~Forr\'{o}}
\affiliation{{Stavropoulos Center for Complex Quantum Matter, Department of Physics and Astronomy, University of Notre Dame, Notre Dame, Indiana 46556, USA}}
\alsoaffiliation{{Laboratory of Physics of Complex Matter, \'{E}cole Polytechnique F\'{e}d\'{e}rale de Lausanne, CH-1015 Lausanne, Switzerland}}

\author{Ferenc~Simon}
\affiliation{{Laboratory of Physics of Complex Matter, \'{E}cole Polytechnique F\'{e}d\'{e}rale de Lausanne, CH-1015 Lausanne, Switzerland}}
\alsoaffiliation{{Department of Physics, Institute of Physics, Budapest University of Technology and Economics, Műegyetem rkp. 3., H-1111 Budapest, Hungary}}
\email{simon.ferenc@ttk.bme.hu}

\begin{document}

\maketitle

\begin{abstract}

Due to their exceptional photovoltaic properties, metal halide perovskites (MHPs) are extensively studied for their potential applications in solar cells. In recent years, the power conversion efficiencies of MHPs-based solar cells rapidly increased from the initial few \% towards more than 25\,\% for single-junction devices. Therefore, also taking into account their low costs and ease of manufacturing, MHPs-based solar cells have become the fastest-advancing photovoltaic technology. In this regard, much of the recent work has been dominated by absorber materials based on methylammonium MHPs, such as MAPbX$_3$, where MA=CH$_3$NH$_3$ and X=Cl, Br and I. Here, we present the results of contactless time-resolved photoconductivity measurements in an exceptionally wide range of temperatures of $4$ to $290\ \text{K}$ that were performed for the various crystalline forms of the three parent MAPbX$_3$, i.e., MAPbCl$_3$, MAPbBr$_3$ and MAPbI$_3$. This approach was made possible by the use of a high quality-factor (Q) microwave resonator, which cooperated with a commercially available microwave bridge equipped with an automatic frequency control (AFC) and a helium gas-flow cryostat.

The structural phase transitions from orthorhombic to tetragonal are found to drastically affect the transient photoconductivity signal and we also observe ultra-long charge carrier recombination times approaching 70 $\mu$s at low temperatures. The difference caused by morphology on the photophysical properties is supported by a marked difference between rapidly cooled (quenched) and slowly cooled samples. The sensitive technique also allowed to observe differences between samples with different morphologies and crystallite sizes.
\end{abstract}

\section*{}

Keywords: Perovskite, Cavity, Photoconductivity, Morphology, Phase transition, Temperature-dependent

\section*{Introduction}
Methylammonium lead halide perovskites (MHPs) emerged as potential candidates to replace conventional silicon-based solar cells due to their impressive photovoltaic properties\cite{efficiency}, relative ease of the crystallization process \cite{doctorblade} and relative insensitivity to defects \cite{defect_insensitivity}. Their additional applications include gas sensorics\cite{gazdetektor}, light-emitting diodes\cite{dioda}, gamma and neutron radiation detectors\cite{rontgen}, as well as their use in space has recently been proposed\cite{spaceLAMI}.

The efficiency of MHP based solar cells is already over 25\,\% \cite{efficiency-uptodate} and is increasing constantly towards the theoretical maximum of single junction solar cells at 33\,\% \cite{perovskitnapelem, perovskitnapelem_ertekezes, perovskitnapelem_nature}. Despite the enormous potential of MHPs, several difficulties have to be overcome before their widespread applications, including degradation issues caused by temperature\cite{nature-temp-degradation}, water or humidity\cite{CH3NH3PbI3:precisestructuralconsequencesofwaterabsorptionatambientconditions} and oxygen\cite{oxigen-hatasa-2}. The lead content of the material poses challenges for the environmentally-friendly disposal of the material. 

MHPs crystallize in the perovskite structure and are orthorhombic at low temperatures (typically below $170\ \text{K}$), followed by a tetragonal and a cubic structure as the temperature increases \cite{dynamic-disorder-in-methylammoniumtrihalogenoplumbates-II-observed-by-milimeter-wave-spectroscopy, review_phasetransitions}. The degrees of freedom of the methylammonium (MA) group is gradually reduced upon lowering the symmetry, induced by the phase transitions at low temperatures. In the orthorhombic phase, the MA cations align in one of two directions which results in localized domains with a built-in potential that can separate charge carrier pairs. \cite{allignment1, allignemnt2, allignment3, allignment4} The phase transitions significantly affect the mobility of charge carriers, however, their role on the photovoltaic properties is yet unclear. The dominant factors that influence the photovoltaic properties of MHPs are a matter of debate, as both the methylammonium cation orientation and ion migration are affected by the temperature\cite{ferro-ion1,ferro-ion2, ferro-ion3, ferro-ion4, ferro-ion5}. 

In principle, temperature dependent photoconductivity studies can resolve these open issues. In particular, time-resolved microwave detected photoconductivity decay\cite{kunst1, kunst2} (TRMCD) measurement is a contactless, non-destructive method, which is widely used in the semiconductor industry to gain information about the wafer quality and purity as well about the charge carrier mobility and photo-excited charge carrier lifetime, $\tau_{\text{c}}$ \cite{Sinton1,sinton2,gyregarami2019ultrafast}.
The latter information is particularly important for photovoltaic applications as the magnitude of $\tau_{\text{c}}$ greatly influences the photovoltaic (PV) cell efficiency: it determines whether the lifetime of light-generated charge carriers is long enough to reach the device electrodes\cite{solyom2} before recombination. A study of $\tau_{\text{c}}$ as a function of temperature can provide information about the charge recombination mechanism and the influence of structural changes. In addition, the amplitude of the TRMCD signal is proportional to the photoconductivity of the material, which in turn characterizes the PV efficiency of the sample. 

We studied the light-induced charge carrier lifetime $\tau_{\text{c}}$ (top) and the transient photoconductivity $\Delta \text{G}$, $\Delta \text{G}$, in methylammonium lead halide perovskites (CH$_3$NH$_3$PbX$_3$, X=I, Br, and Cl; denoted as MAPbX$_3$ in the following) in the $4-270\ \text{K}$ temperature range. We find clear evidence that the orthorhombic to tetragonal phase transition affects both $\tau_{\text{c}}$ and $\Delta \text{G}$, while the influence of the tetragonal to cubic transition on the photoconductivity is less pronounced. 
The longest charge carrier lifetime of 68 $\mu$s was observed in MAPbBr$_3$ at $10\ \text{K}$. We also find evidence that the sample crystallization procedure, as well as the cooling protocol, have observable roles. These results shed light on the interplay of photovoltaic-relevant properties and the structure in these materials, whose understanding can lead to further improvements in the PV efficiency. 

\section*{Results and discussion}

\begin{figure}[!h]
	\centering
	\includegraphics[width=0.98\linewidth]{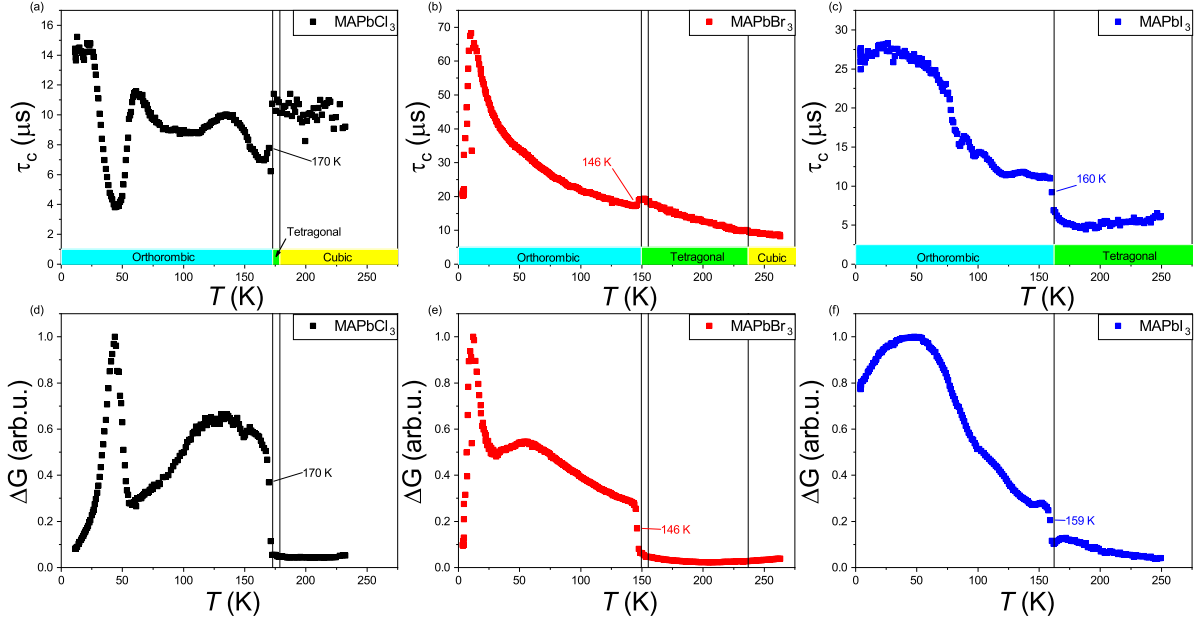}
	\caption{Temperature-evolution of charge carrier lifetime $\tau_{\text{c}}$ (top) and photo-conductance $\Delta \text{G}$ (bottom) for crystals of three parent MHPs, $i.e.$, MAPbCl$_3$, MAPbBr$_3$ and MAPbI$_3$. A marked change in both the recombination time and photoconductivity can be observed at the orthorhombic-to-tetragonal phase transition, which is indicated by vertical lines. The materials show ultra-long charge carrier recombination times at low temperatures with the maximal value exceeding $68\ \mu\text{s}$ for MAPbBr$_3$. The measurements were made with a Q-switched nanosecond pulsed laser operating at $450\ \text{nm}$. The data points were recorded while raising the temperature with the TRMCD method}
	\label{Fig1_Tau_Delta_G}
\end{figure}

The result of the temperature-dependent microwave photoconductivity decay measurements for crystals of all three parent MHPs down to $4 \ \text{K}$ are shown in Fig. \ref{Fig1_Tau_Delta_G}. We show the temperature dependence of two properties: the charge carrier lifetime, $\tau_{\text{c}}$ and the photoconductivity, $\Delta \text{G}$. The former is determined from the time constant of the photoconductivity decay curves and the latter from its initial amplitude, as described in the Methods section. All photoconductivity decay curves can be described with a single exponent, indicating the presence of a dominant recombination channel.

The orthorhombic to tetragonal phase transition is found to have a relatively smaller effect on $\tau_{\text{c}}$ in the case of MAPbBr$_3$, and more promminent in the case of MAPbCl$_3$ and MAPbI$_3$ as shown in Fig. \ref{Fig1_Tau_Delta_G}. More interestingly, the orthorhombic-to-tetragonal phase transition has a well-pronounced impact on the photoconductivity, which was observed to increase by a factor of $\sim 10$, $\sim 6$ and $\sim 2.5$, for MAPbCl$_3$, MAPbBr$_3$, and MAPbI$_3$, respectively. The first order nature of the phase transition also manifests in a hysteretic behavior on the photoconductivity, which is presented in detail in the Methods section. This hysteretic behavior caused by the temporal coexistence of both phases around the phase transition was previously observed by Osherov $et\ al$ \cite{structural_hyst}. There is also a generic trend that the photoconductivity reaches a maximum in the orthorhomic phase which is followed by a decrease towards lower temperatures. The temperature for the maximum in the photoconductivity also shows a halide dependence: between the MAPbCl$_3$ and the MAPbBr$_3$ the observed narrow peak with the maxima shifted towards lower temperatures. There might be a similar peak for the MAPbI$_3$ sample below our measurement range, if there is a trend of decreasing peak temperature as the atomic size of halogens increases. The temperature of the peaks and the observed lifetimes are summarized in Table \ref{maxrecomb}.

The influence of the structure on the photoconducting properties was observed in Ref. \cite{built-in-E} for MAPbI$_3$ and MAPbBr$_3$ which was associated to the ferroelectric ordering of the MA groups, that results in a built-in electric field which assists both the charge separation (thus increasing the photoconductivity), and also increases the charge carrier lifetime \cite{ferro-ion1,ferro-ion2}. The appearance of the maximum in $\Delta \text{G}$ at low temperatures, which is followed by a decreasing photoconductivity towards lower temperatures is tentatively assigned to a yet unobserved but predicted antiferroelectric ordering of the methylammonium groups \cite{Leguy_MA_ordering}. Such an ordering is expected to cancel the positive effect of the ferroelectric ordering thus reducing the observed photoconductivity.

\begin{table}[h!]
	\begin{tabular}{c|c|c|c}
		& $T_c\ \text{(K)}$ & $\tau_{\text{c}} \ \text{(}\mu\text{s)}$ &$T_{\Delta \text{G}}\ \text{K}$ \\ \hline
		MAPbCl$_3$ & $13$ & $15.2$ & $44$    \\ \hline
		MAPbBr$_3$ & $10$ & $68.3$ & $12$   \\ \hline
		MAPbI$_3$ & $26$ & $28.2$ & $48$    
	\end{tabular}
	\caption{Maximal values of the charge carrier lifetime, $\tau_{\text{c}}$ for the measured methylammonium lead halide perovskites, the temperature where the value is attained, $T_c$ and the temperature of the maximal value of photoconductivity, $T_{\Delta \text{G}}$.}
	\label{maxrecomb}
\end{table}

The tetragonal to cubic transition has no visible effect on either the charge carrier lifetime or the photoconductivity. This supports the earlier assignment that the MA ordering has a pronounced effect on the photovoltaic properties, which only occurs in the orthorhombic state. For crystals of MAPbBr$_3$ and MAPbI$_3$, we observe a gradual increase in $\tau_{\text{c}}$ with decreasing temperature, which can be attributed to gradually increasing constraints on orientational dynamics of MA cations upon transition to the orthorhombic phase. It is assumed that MA groups lose their orientational degree of freedom and only two possible orientations remain allowed\cite{allignment4}. This loss of movement is drastically different from the one observed at the cubic-tetragonal phase transition thus the enormous difference in the effect the change has on the recombination process. Polarized domains of unidirectional organic cations emerge inside the material resulting in a built-in local electric field that propels the generated charge carriers away from each other this way elevating the recombination time\cite{built-in-E}.

We observe ultra-long charge carrier lifetimes in all methylammonium lead halide perovskites at low temperatures, with the values summarized in Table \ref{maxrecomb}. The observed values are all larger than those found in earlier TRMCD studies \cite{TRMCD_10us,TRMC_SS,TRMC_cavity,TRMC_cavity2,TRMC_thin_film,TRMC_thin_film2,TRMC_transmission}, which may be related to an improved sample quality in the present study. 

We also address the slight drop in the charge carrier lifetime at the phase transition from tetragonal to orthorhombic phase for the MAPbBr$_3$ and MAPbI$_3$ samples. We believe that it is due to the coexistence of orthorhombic and tetragonal domains right below the phase transition. Charge carriers are scattered at the domain boundaries thus lowering their lifetime. By further decreasing the temperature the whole crystal becomes orthorhombic and the domains disappear. The coexistence of a small number of domains in the tetragonal phase in the regime of the orthorhombic phase was observed previously \cite{temp-mobility, temp-PL}. Interestingly the $\text{MAPbCl}_3$ sample shows the opposite change for the lifetime, as $\tau_{\text{c}}$ increases in the orthorhombic phase compared to the value observed in the tetragonal one.

\begin{figure}[!h]
	\centering
	\includegraphics[width=0.98\linewidth]{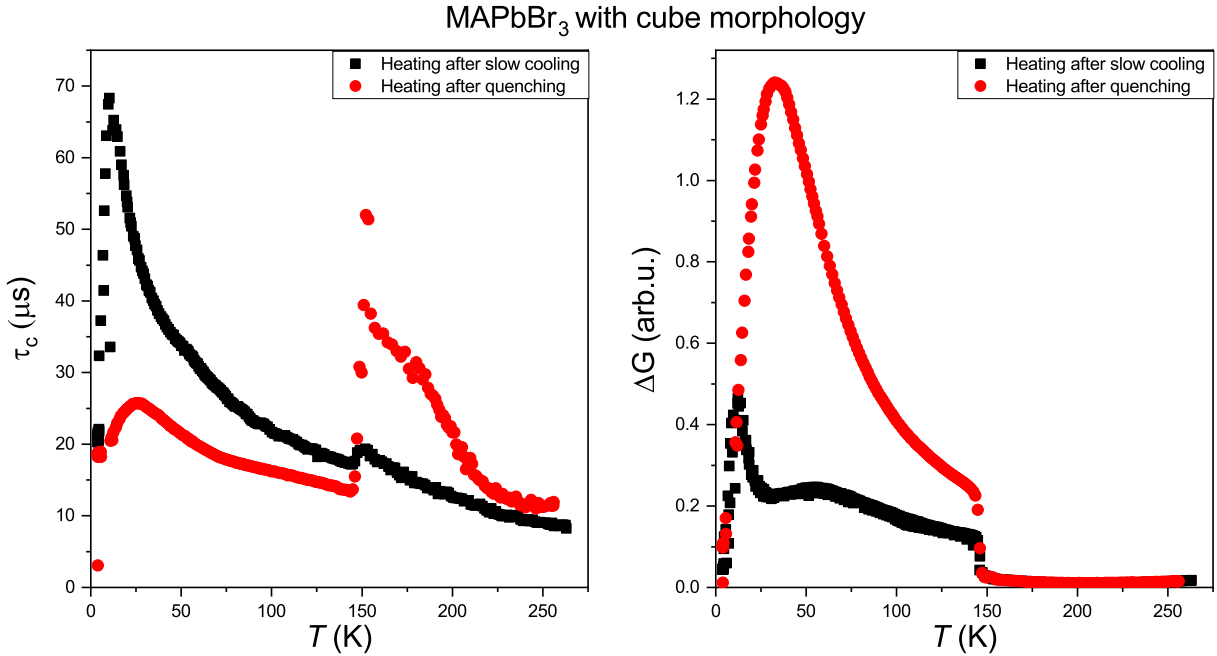}
	\caption{The charge carrier lifetime, $\tau_{\text{c}}$, and the photoconductivity, $\Delta \text{G}$, in quenched and non-quenched MAPbBr$_3$ cubic crystals as measured by TRMCD during slow gradual heating. In the orthorhombic phase, $\tau_{\text{c}}$ is significantly shorter for the quenched crystal than for its non-quenched counterpart (left panel). In contrast, $\Delta \text{G}$ shows the opposite trend, reaching much higher values for the quenched crystal (right panel). Note also a marked jump in both measured properties, $\tau_{\text{c}}$ and $\Delta \text{G}$, at the orthorhombic-to-tetragonal phase transition. In particular, upon this phase transition, the initial large overshoot in $\tau_{\text{c}}$ for the quenched crystal quickly falls down to the level observed for its non-quenched counterpart (left panel).}
	\label{Fig2_Quench}
\end{figure}

The interplay between crystal structure and electronic properties can be suitably studied with varying the cooling speed. An extreme case of this investigation is when the sample is rapidly cooled or \emph{quenched} from room temperature to $4\ \text{K}$. The construction of our cryostat (Oxford Instr. Sci.Div., Model ESR900) allows this by precooling it to $4\ \text{K}$ and then immersing the sample holder, which contains the sample into it. Given the small sample volume, we estimate that it thermalizes within less than a few seconds. A similar protocol helped previously to stabilize otherwise thermodynamically unstable alkali fulleride structures \cite{lappas_quench}.

In Fig. \ref{Fig2_Quench}. the effect of quenching on a MAPbBr$_3$ sample is presented. First, the sample was quenched from room temperature to $4\ \text{K}$ and then subjected to a slow gradual heating, during which the TRMCD measurements were performed to get insight into the above-mentioned properties, that is, $\tau_{\text{c}}$ and $\Delta \text{G}$. Then, the sample was cooled down once more, while it remained in place in the cryostat and warmed up again slowly. We compare the data for the warming up protocols. Consecutive slowly cooled and warmed up runs reproduce the previous slow cooling results. 

The photoconductivity increases below the tetragonal to orthorhombic phase transition for both types of experiments, however, the increase of $\Delta \text{G}$ is significantly larger for the quenched sample. It is important to note that the two set of $\Delta \text{G}$ data were not normalized as the sample was not removed from the cavity between the two measurements. While $\Delta \text{G}$ is larger for the quenched sample, the corresponding charge carrier lifetime is smaller and the low temperature peak is less pronounced. We believe that both effects can be assigned to a lower structural order in the orthorhombic phase, probably in the form of smaller ferroelectrically oriented structural domains. Small domain sizes give rise to increased scattering thus reducing $\tau_{\text{c}}$. As we observed in the case of samples with varying morphology, the lowered symmetry increases the photoconductivity, which explains the behavior in the quenched sample. 

Another compelling effect is the sudden upright jump in $\tau_{\text{c}}$ for the quenched sample immediately when it is warmed above the orthorhombic-to-tetragonal phase transition. Again, this effect is robust and it was observed for several quenching runs. Although, we do not have a consistent explanation for this effect, we propose that a possible coexistence of tetragonal and orthorhombic domains may play a role, which merits further studies, for example by means of the crystallographic structural investigation in the quenched sample.

\begin{figure}[!h]
	\centering

	\includegraphics[width=0.98\linewidth]{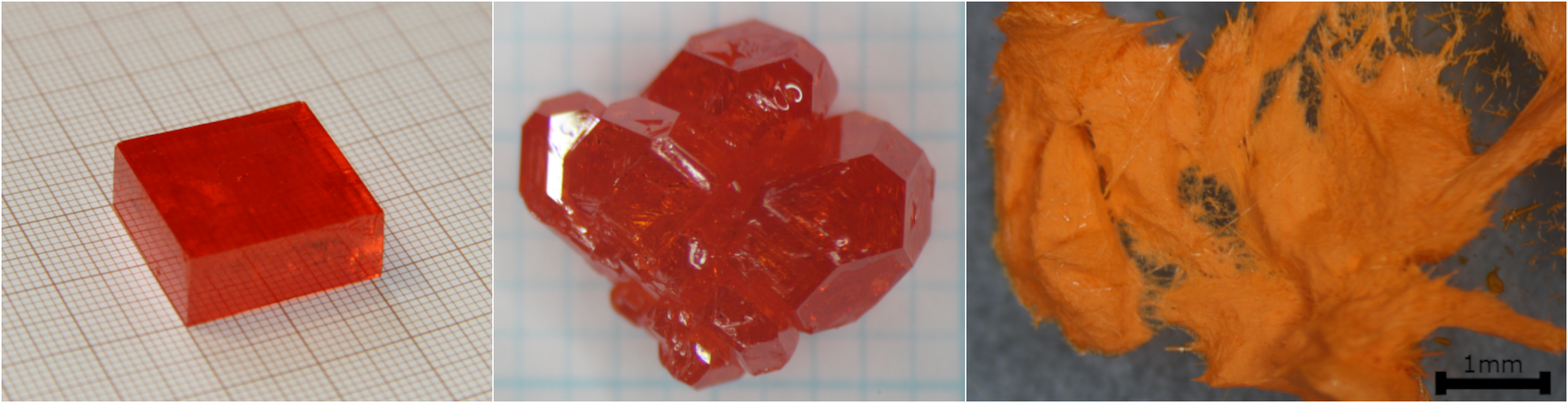}
	\includegraphics[width=0.98\linewidth]{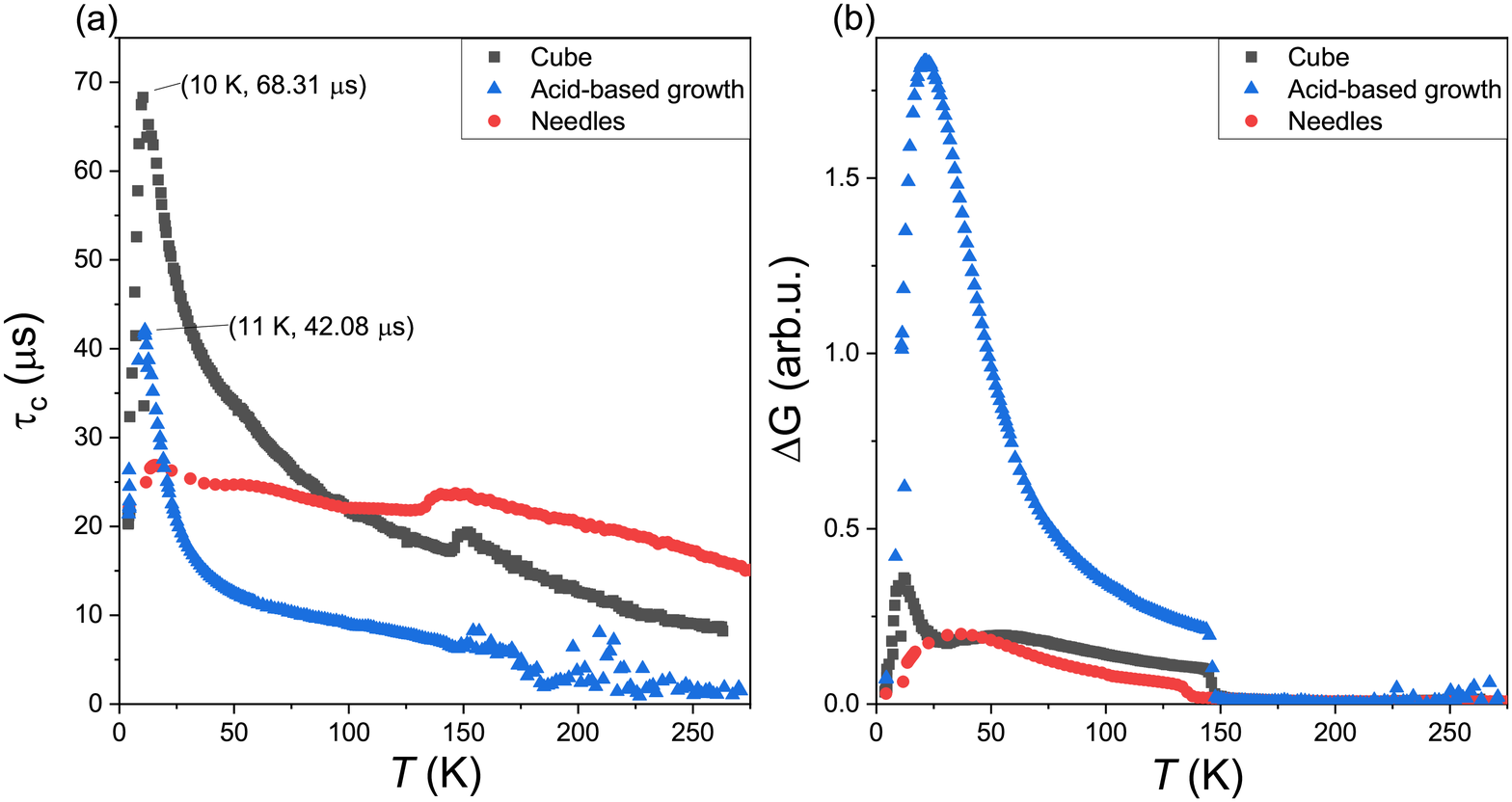}
	\caption{Impact of the crystalline morphology of MAPbBr$_3$ on the charge carrier lifetime, $\tau_{\text{c}}$, and the photoconductivity, $\Delta \text{G}$, as probed by TRMCD. Top panel, from left to right: photographs of cubic, amorf (acid-based crystal growth), and needle-shaped samples (the scale is indicated by the raster grid lines on corresponding paper sheets or by a scalebar). The cubic samples reveal the longest charge carrier lifetimes, whereas the photoconductivity is largest for the acid-based samples. The $\Delta \text{G}$ data is normalized by the respective values above the orthorhombic-to-tetragonal phase transition.}
	\label{Fig3_morphology_withsample}
\end{figure}

In Fig. \ref{Fig3_morphology_withsample}. we show the difference in the charge carrier lifetime and photoconductivity for three different morphologies of MAPbBr$_3$ samples. The top panel in the figure shows photographs for these morphologies: a cube-shaped crystal, a less crystalline sample made using a different, acid-based growth process, and a sample consisting of needle-shaped crystallites. The synthesis routes for these sample types are detailed in the Methods section. 

The sample properties display characteristic differences for the measured parameters: the charge carrier lifetime has a tendency to be longest for the sample with the best crystallinity (the cubic sample), whereas the photoconductivity is largest below the tetragonal-orthorhombic transition for the acid-based sample. We note that the latter statement relies on a normalization of the $\Delta \text{G}$ data in the tetragonal phase (as shown in Fig. \ref{Fig3_morphology_withsample}.) since the measured photoconductivity in our method does not provide absolute values and depend on the sample amount. This normalization also attempts to eliminate artefacts such as a higher absorption cross section of the needle-shaped samples as compared to the cubic ones (e.g. due to the index of refraction discontinuity on the sample surface) as we expect that these effects would be the same irrespective when the sample undergoes the structural phase transition. We verified that the observed properties are reproducible for several different samples from various batches, produced with the same synthesis methods.

The impact of the crystalline morphology on the temperature dependencies of $\tau_{\text{c}}$ and $\Delta \text{G}$ is also worth discussing. The cube-shaped and the acid-based samples have similar temperature dependencies, since both show the longest charge carrier lifetimes at low temperatures. Likewise, for these both crystalline versions of MAPbBr$_3$, $\Delta \text{G}$ reveals a pronounced jump at the orthorhombic-to-tetragonal phase transition ($\sim148-154\ \text{K}$), while for the needle-shaped sample slight step changes in $\tau_{\text{c}}$ and $\Delta \text{G}$ values are shifted towards lower temperatures by $\sim15\ \text{K}$ (to $\sim135\ \text{K}$). Moreover, compared to cube-shaped and the acid-based samples, the needle-shaped sample shows significantly different temperature dependencies of $\tau_{\text{c}}$ and $\Delta \text{G}$. In particular, the temperature dependence of $\tau_{\text{c}}$ for the needle-shaped sample is almost featureless as compared to the other two samples.

These observations support the previous generic findings from the quenching experiments: lowering the crystal symmetry (such as in the orthorhombic phase) and inducing disorder either by quenching or a sample witch consists of multiple crystallites results in a shorter charge carrier lifetime, probably due to the presence of grain boundaries which act as recombination centers. However, at the same time, the induced disorder caused by quenching has a positive effect on the photoconductivity. On the other hand, a fully disordered sample, such as that observed for the needle-shaped ones, again result in reduced photoconductivity. We note that our result does not provide information about the recombination mechanism. It is known for standard photovoltaic materials (such as e.g., Si) that three competing mechanisms contribute to this: impurity assisted recombination (also known as Shockley-Read-Hall), radiative recombination, Auger process.

The efficiency of a photovoltaic cell depends on both of these factors: the charge carrier lifetime, which determines whether the electron-hole pair can reach the respective electrodes, and the photoconductivity which is directly proportional to the number of generated charge carrier pairs. Our observations suggest that the optimal sample morphology has to be determined with this trade-off in mind. E.g., for a thin film photovoltaic cell, where a short charge carrier lifetime limits little the performance, a sample type with low crystallinity is preferred, whereas for a cell where a large current capability or mechanical robustness is required, the cubic sample type is the preferable.

\section*{Conclusions}

To summarize, we studied the photoconductivity and the charge carrier lifetime of photoinduced charge carriers using a microwave reflectometric method down to liquid helium temperatures for the three methylammonium lead halide perovskites (CH$_3$NH$_3$PbX$_3$, X=I, Br, and Cl). An ultra-long charge carrier recombination time in the orthorhombic phase for each methylammonium lead halide perovskite was observed. The structural changes have a clear effect on both of these parameters, namely the low temperature orthorhombic phase, where a ferroelectric ordering of the methylammonium cation groups is present, shows improved properties, which may be associated with improved photovoltaic efficiencies. These findings are corroborated by rapidly cooled studies which indicates structural disorder and also on samples with varying morphologies. Our findings shed light on the delicate interplay between structure and PV efficiencies in these materials which may help to construct a variant where the improved properties are retained at room temperature.

\section*{Methods and tools}

\subsection{Sample preparation}

Acid growth: Crystals of the methylammonium lead trihalide (CH$_3$NH$_3$PbX$_3$, X=Br, Cl, I) were synthesized by solution growth. The $3.3\ \text{mmol}$ lead (II) acetate trihidrate (Pb(ac)$_2\times$ $3\text{H}_2\text{O}$, $>99.9$ \%) was reacted with $6\ \text{ml}$ saturated HX solution (saturated HX solution in $\text{H}_2\text{O}$). The formed $\text{PbX}_2$ precipitate is stable in the acidic solution. The respective amount ($3.30\ \text{mmol}$) methylamine ($\text{CH}_3\text{NH}_2$) solution ($40\ \text{wt\%}$ in $\text{H}_2\text{O}$) was pipetted into the $5\ ^\circ \text{C}$ ice-cooled solutions of $\text{PbX}_2$. The cold solution avoids the evaporation of methylamine during the exothermic reaction. The microcrystallites of $\text{CH}_3\text{NH}_3\text{PbX}_3$ were formed immediately and settled down at the bottom of the vessel. In a temperature gradient of $15\ ^{\circ}\text{C}$ in the acidic media, large ($1-8\ \text{mm}$) perovskite single crystals grew on the cold side of the vessel.
 Inverse temperature growth: $\text{MAPbX}_3$ single crystals were synthesised using the inverse temperature crystallisation method from its saturated solution in dimethylformamide (DMF). $\text{MAPbBr}_3$ was dissolved in DMF at room temperature. Crystal growth was initialized by increasing the temperature of the solution from room temperature to $50\ ^\circ \text{C}$ with a heating rate of $0.5\ ^\circ \text{C}$ per hour as depicted. Nice cubic-like, centimeter-sized single crystals can be harvested after only a couple of hours of crystallogenesis. Besides the bromide version, $\text{MAPbI}_3$ and $\text{MAPbCl}_3$ can be grown using the identical method, with $\gamma$-butyrolactone (GBL) and $50\ \text{V/V\%}$ DMF / $50\ \text{V/V\%}$ dimethyl sulfoxide (DMSO) in the case of perovskites containing iodide and chloride, respectively.

Needle shaped $\text{MAPbBr}_3$ crystals: Needle shaped perovskite crystals were grown by cooling down the room temperature saturated perovskite solution in DMF to $-15 ^\circ \text{C}$, where needle shape crystals were formed. The harvest of these crystals was done at low temperature and the DMF solvent was removed by freeze-drying.

\subsection{Microwave-detected photoconductivity measurement}

In this study we present a simple and so far not implemented approach to the contactless probing of photoconductivity with microwaves for the crystalline forms of MHPs compounds, namely MAPbI$_3$, MAPbBr$_3$ and MAPbCl$_3$. In particular, our experimental setup for contactless measurement of light induced changes in conductivity is based on commercially-available building blocks, which can be found in typical electron spin resonance spectrometers. Thanks to this, we were able to take advantage of two key elements, which have been optimized to carry out sensitive measurements of the paramagnetic susceptibility. The sensitivity of the measurement is ensured by placing the sample in a microwave resonator with high Q-factor (loaded Q-factor value, Q$_L$ $\approx6550$) and by employing a frequency-stabilized microwave source, which, thanks to automatic frequency control, locks the resonant frequency of the microwave generator to that of the microwave resonator. This solution allows for the sensitive measurement of small changes in the magnetic susceptibility occurring in the sample under the influence of a changing magnetic field. Furthermore, the automatic frequency control takes care if the resonator frequency shift this way eliminating the dispersive component resulting in the detection of the absorptive component only. We utilized this characteristic property of the microwave bridge and cavity of a standard electron spin resonance spectrometer to separate the dispersive and absorptive components of photoconductance. The novelty of the experimental setup lies in the combination of the type of microwave cavity, the placement of the sample in relation to the microwave field and the method of bringing the exciting laser light to the samples.

\begin{figure}[!h]
	\centering
	\includegraphics[width=0.73\textwidth]{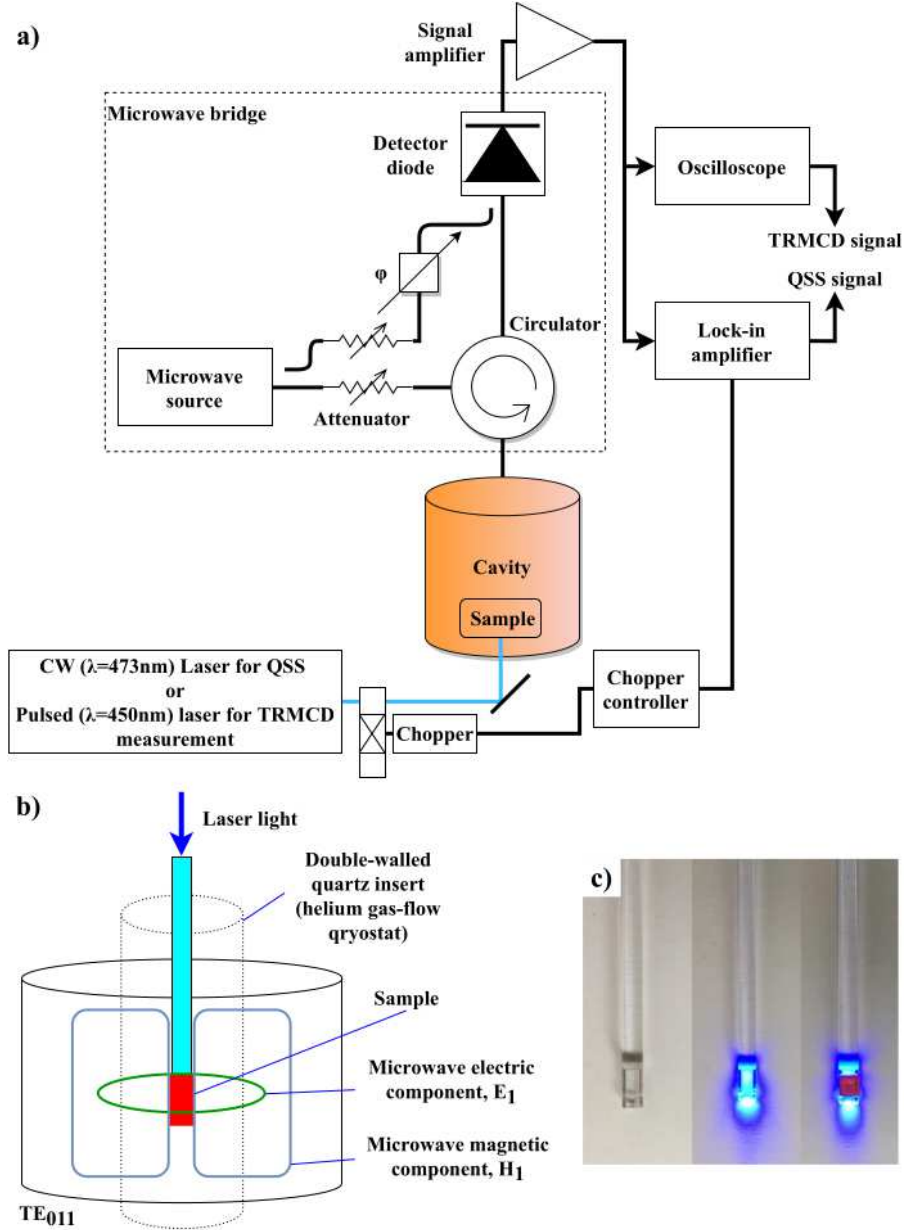}
	\caption{The experimental setup used in this work for microwave photoconductivity measurements. (a) Block diagram of the system including its modifications corresponding to TRMCD and quasi-steady state measurements. (b) Schematic representation of the laser light delivery to a sample placed in the center of a cylindrical resonator operating in TE$_{011}$ mode. (c) Illustration of the light delivery to the incission in the $4-\text{mm}$ diameter quartz rod where the sample is mounted. Photos of the bottom part of the quartz rod taken: without illumination, under illumination and under illumination and with the MAPbBr$_3$ crystal attached are shown in the left, middle, and right panels respectively. For the purpose of visualization of the light delivery to the sample, the photos were taken under illumination with a fiber-coupled CW LED light source ($\lambda=470\ \text{nm}$, ADSresonances Sàrl).}
	\label{block}
\end{figure}

The diagram of the experimental system that was used in this work for the microwave measurement of photoconductivity in parent MHPs is shown in Fig. \ref{block}. The sensitive detection is based on using the microwave bridge of a commercial electron spin resonance spectrometer (Bruker Elexsys E500). The use of a microwave cavity amplifies the effect of light induced changes in the photoconductivity as shown in Ref. \cite{Janes1998,gyregarami2019ultrafast, Savenije-cavitysetup, savenije_MHPTRMC}, whereas its finite bandwidth only slightly limits the detectable transient signals to $\tau_{\text{tr}}=2Q/\omega_0$. Here, $Q\approx 6550$ is the quality factor of the loaded cavity and $\omega_0\approx 2\pi \cdot 9.5\,\text{GHz}$, which gives $\tau_{\text{tr}}\approx 220\,\text{ns}$. This also means that we inevitably miss any faster dynamic processes. However, the known typical recombination times are usually above $1\ \mu\text{s}$. Given that the targeted decay lifetimes are beyond the $\mu\text{s}$ timescale, this does not significantly affect the measurement. We note that the microwave output of the Bruker broad-band signal preamplifier ($20\ \text{Hz-}6.5\ \text{MHz}$, model ER 047-PH) follows a low-pass filter characteristics with a bandwidth of 6 MHz, which also gives a limiting transient time of $\tau_{\text{tr}}=1/2\pi\cdot 6\text{MHz}=25\,\text{ns}$, again not affecting our measurements.

The sample is positioned at the center of the cylindrical TE$_{011}$ microwave cavity (BioSpin high-Q cavity, Model ER 4122SHQE) inside a double-walled quartz insert, which sustains the helium gas flow, thus allowing to control the sample temperature between $4\ \text{K}$ and $270\ \text{K}$. The samples are in the center of the TE011 cavity where the microwave electric field has a node, and the magnetic field has its maximum. However, the exact center is surrounded by a circularly rotating electric field which induces currents in the sample whose loss is detected.\cite{csosz_cavity} The automatic frequency control (AFC) of the commercial CW X-band microwave bridge (Model ER049X from Bruker BioSpin) locks the frequency of the microwave source (a Gunn diode) on the actual resonance frequency of the cavity. Thanks to this, it is possible to relatively easily isolate the real (dispersive) and the imaginary (absorptive) parts of the changes in the overall sample susceptibility \cite{cavity1,cavity2}. Even when the temperature affects the latter. The automatic frequency control of the microwave bridge has a frequency of $76.8\ \text{kHz}$. Such an experimental setup significantly facilitates the direct measurement of the changes in the Q-factor of the microwave cavity, which are related to the light-induced changes in the sample conductivity.

In order to be able to bring light to the sample positioned at the center of a high-Q resonant cavity (otherwise devoid of illumination slots), we used a $25\ \text{cm}$ long section of a clear fused quartz rod with diameter of $4.0\ \text{mm}$ (Spectrosil 2000 from Heraeus Quarzglas GmbH $\&$ Co., Kleinostheim, Germany), which, due to its high refraction index (of $\sim 1.46$ at $470\ \text{nm}$), served as an efficient light-guide. The laser beam light was focused on the top edge of the quartz rod, while the sample was attached (glued) at its bottom notch. This method of excitation is similar to the one presented by Mantulnikovs $et\ al$ \cite{mantulnikovs2020light}.

We used two different measurement setups, which served for quasi-steady state (QSS) and time-resolved (TR) microwave photoconductivity measurements. In the case of the QSS approach, the beam of the laser light ($473\ \text{nm}$, $50\ \text{mW}$) from a CW fiber-coupled laser (Model MBL-III-473, from CNI Lasers Ltd.) was modulated with a mechanical optical chopper (Model SR540 from Stanford Research Systems, Inc.). To perform TR measurements, we used a Q-switched nanosecond pulsed laser-diode system (Model NPL45C from Thorlabs Inc.) which operated at $450\ \text{nm}$, with a pulse duration of $6-129\ \text{ns}$ and a pulse energy of $200\ \text{nJ}$. The wavelength of the Q-switched laser is close to the optical absorption edge in the case of MAPbCl$_3$ \cite{absorptions_lami, lami_absorption} while according to our results the excitation of the sample was achieved sufficiently nonetheless. 

\begin{figure}[!h]
    \centering
    \includegraphics[width=0.48\linewidth]{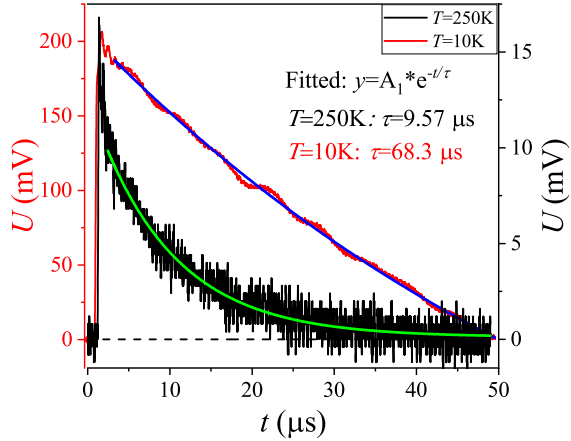}
    \caption{Example time-traces of the measurement fitted with a single exponent. The two traces were recorded during the same temperature sweep and with the same excitation. The time-traces were collected during the measurement presented in Fig. \ref{Fig2_Quench}.}
    \label{timetrace}
\end{figure}

The spectrometer output is fed to a lock-in amplifier (Stanford Research Systems SR830) for the QSS and to an oscilloscope for the TR measurement. A representative time-trace measurement is also shown in Fig. \ref{timetrace}. The clear advantage of the lock-in measurement is its higher sensitivity due to the well controlled measurement bandwidth, however its disadvantage is that the measured signal is the product of the photoconductivity and the charge carrier lifetime thus the two quantities cannot be separated.

\begin{figure}[!h]
	\centering
	\includegraphics[width=0.48\linewidth]{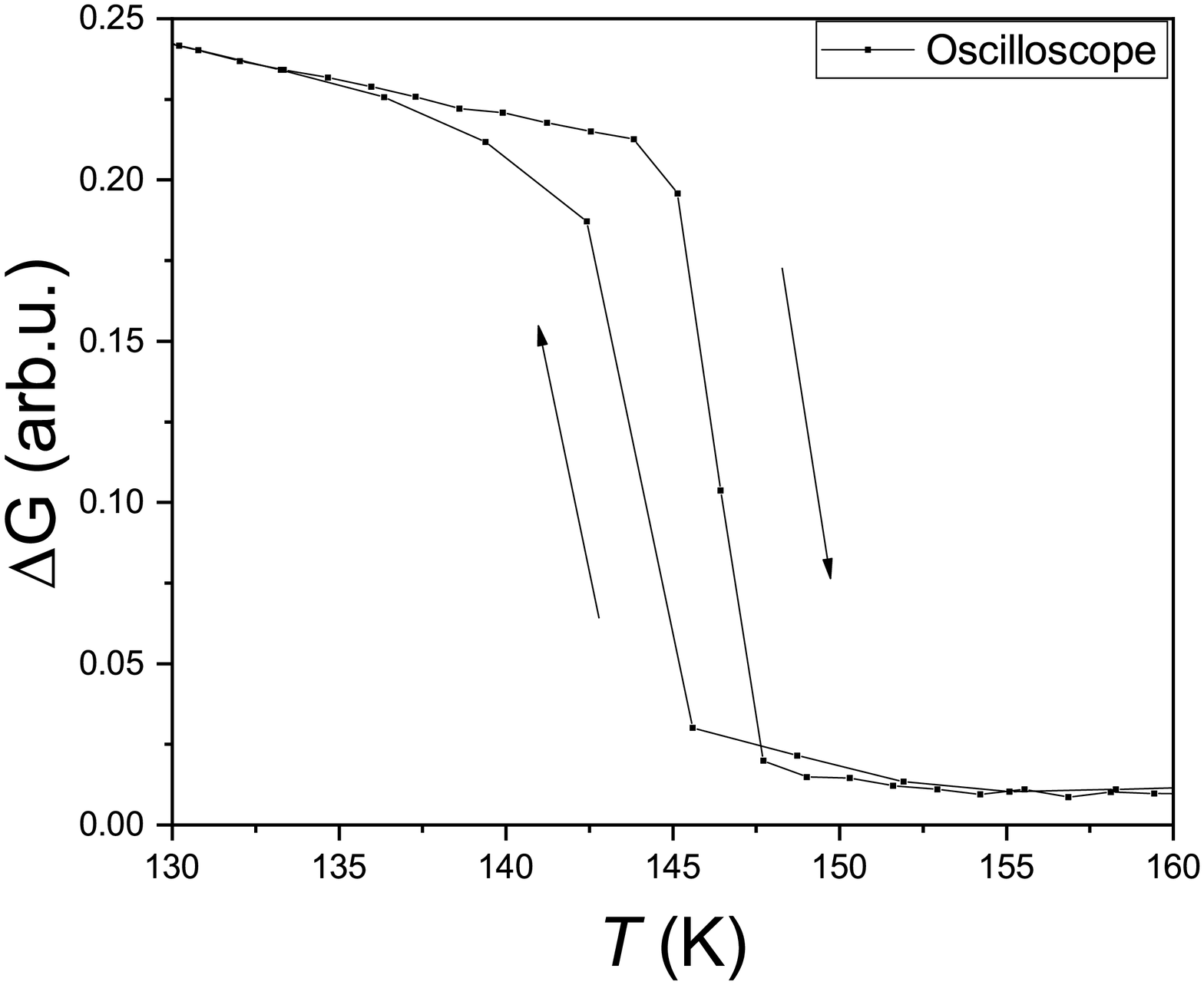}
	\includegraphics[width=0.48\linewidth]{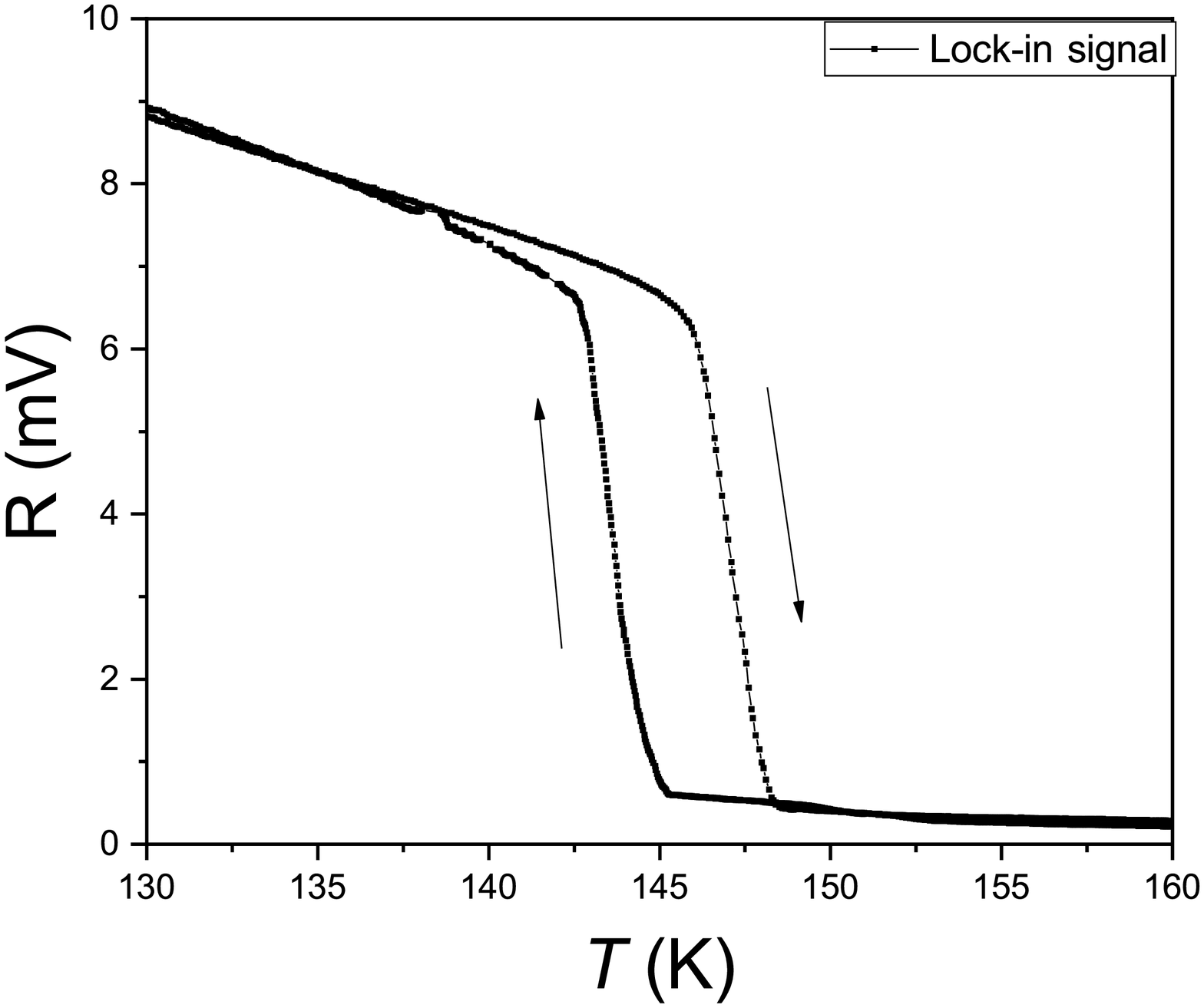}
	\caption{Hysteresis of the microwave photoconductivity signal at the tetragonal-orthorhombic phase transition for the time-resolved (left) and the quasi-steady-state (right) measurements for CH$_3$NH$_3$PbBr$_3$. Note the larger number of points for the QSS measurement as a result of the noise-free lock-in technique. }
	\label{hister}
\end{figure}

The differences and similarities between the two types of measurements are demonstrated in Fig. \ref{hister}., where the photoconductivity is presented near the tetragonal-orthorhombic phase transition for CH$_3$NH$_3$PbBr$_3$. The hysteresis in the photoconductivity is much better observed for the QSS technique as more points can be obtained for the same measurement time and warming protocol due to the higher sensitivity of the lock-in detection technique.

In the following, we show that the relationship between the spectrometer output and the sample photoconductivity is linear. In high frequency measurements, the properties of materials are conveniently described by the concept of surface impedance at a finite electromagnetic frequency of $\omega$\cite{pozar, LFChen}:

\begin{equation}
	Z_{\text{s}}=\sqrt{\frac{\mathrm{i}\omega\mu}{\sigma+\mathrm{i}\omega\epsilon}}
\end{equation}
which contains the effects of magnetism (through the permeability, $\mu$), dielectrics (through the permittivity, $\epsilon$) and conductivity (through $\sigma$). 

A sample with surface impedance $Z_{\text{s}}$ affects both the resonance frequency, $f_0$ and the quality factor, $Q$ of the cavity through\cite{Klein, landau2013electrodynamics}:

\begin{equation}
	\frac{\Delta f}{f_0}+\mathrm{i}\Delta \left(\frac{1}{2Q}\right)=\mathrm{i}\nu Z_{\text{s}} 
\end{equation}
where $\Delta f$ is the resonance frequency shift of the empty cavity and $\nu$ is the so-called filling factor, roughly equal to the sample volume per the cavity volume ratio. This expression shows that both the resonance frequency shift and the $Q$ are directly connected to the surface impedance of the sample. The microwave bridge of our spectrometer was set up to detect the change in the $Q$ factor only, thus changes in the imaginary part of $Z_{\text{s}}$ is measured \cite{PooleBook}. The instrument output is linear in $\Delta (\frac{1}{2Q})$. 

The illumination causes an extra conductivity of the sample, $\Delta \sigma$, which also affects the sample surface impedance as:

\begin{equation}
	Z_{\text{s}}(\sigma+\Delta\sigma)=\sqrt{\frac{\mathrm{i}\omega\mu}{\mathrm{i}\omega\epsilon+\sigma+\Delta\sigma}}
\end{equation}

Given that the MHP materials are poor conductors without illumination, $\omega\epsilon\gg \sigma$ thus a series expansion with the small parameter $\Delta \sigma$ yields:
\begin{equation}
Z_{\text{s}}(\sigma+\Delta\sigma)\approx Z_{\text{s}}(\sigma)\cdot \left(1- \frac{\Delta\sigma}{2\left(\mathrm{i}\omega\epsilon+\sigma\right)} \right).
\end{equation}
Therefore the excess spectrometer output due to the illumination is linear in $\Delta \sigma$, as expected.

\subsection{Charge carrier generation}

The change in the number of charge carrier pairs in a given sample is represented by the so called rate equations:

\begin{equation}
    \frac{dn}{dt}=-\frac{n}{\tau}+\text{G}_{\text{photo}}
\end{equation}

where $\tau$ is the trap limited charge carrier lifetime, $\text{G}_{\text{photo}}$ is the carrier generation rate and $n$ is the carrier concentration. The differential equation can be solved in different regimes, including the start of a laser pulse, a quasi steady state, and the relaxation of the generated charge carriers following a laser pulse. The general solution for this equation gives the following time-dependent charge carrier concentration:

\begin{equation}
    n(t)=c  \cdot \text{e}^{-t/\tau}+\text{G}_{\text{photo}} \cdot \tau
\end{equation}

, where $c$ is constant determined by the initial conditions. In the case where the laser pulse is just beginning, increasing the charge carrier concentration the $n(0)=0$ assumption can be made, resulting in the following solution:

\begin{equation}
    n(t)=\text{G}_{\text{photo}} \cdot \tau(1-\text{e}^{-t/\tau})
\end{equation}

Since the carrier lifetime is orders of magnitude longer than the full width at half maximum of the laser pulse the recombination process will not be significant during the time frame. This can be shown by calculating the Taylor series around $t=0$, which gives $e^{-t/\tau}\approx 1-\frac{t}{\tau}+\frac{t^2}{2\tau^2}-....$. This leads to the charge carrier concentration during the laser pulse to be proportional only to the carrier generation rate:

\begin{equation}
    n(t)=\text{G}_{\text{photo}}\cdot t
\end{equation}

By assuming a quasi steady state ($n'=0$) for the same differential equation yields:

\begin{equation}
    n=\text{G}_{\text{photo}} \cdot \tau
\end{equation}

If one is looking at the relaxation of the charge carriers the charge carrier generation should be assumed to be zero ($\text{G}_{\text{photo}}=0$). This leads to a solution that consists of only an exponential component:

\begin{equation}
    n(t)=n_0 \cdot \text{e}^{-t/\tau}
\end{equation}

, where $n_0$ is the initial population.

\section*{Funding Sources}
This work was supported by the Hungarian National Research, Development and Innovation Office (NKFIH) Grant K137852, TKP2021-EGA-02, TKP2021-NVA-02, and the Quantum Information National Laboratory sponsored by the Ministry of Innovation and Technology via the NKFIH. This work was supported by the KDP-2021 Program of the Ministry for Innovation and Technology from the source of the National Research, Development and Innovation Fund.

\section*{Author contributions}
Samples were prepared by MK under the guidance of LF. Magnetic resonance and photoluminescence characterization was provided by AS. AB, SK, BGM, and FS performed the photoconductivity measurements and the data were analyzed by AB. All Authors contributed to writing the paper.


\begin{mcitethebibliography}{59}
	\providecommand*\natexlab[1]{#1}
	\providecommand*\mciteSetBstSublistMode[1]{}
	\providecommand*\mciteSetBstMaxWidthForm[2]{}
	\providecommand*\mciteBstWouldAddEndPuncttrue
	{\def\EndOfBibitem{\unskip.}}
	\providecommand*\mciteBstWouldAddEndPunctfalse
	{\let\EndOfBibitem\relax}
	\providecommand*\mciteSetBstMidEndSepPunct[3]{}
	\providecommand*\mciteSetBstSublistLabelBeginEnd[3]{}
	\providecommand*\EndOfBibitem{}
	\mciteSetBstSublistMode{f}
	\mciteSetBstMaxWidthForm{subitem}{(\alph{mcitesubitemcount})}
	\mciteSetBstSublistLabelBeginEnd
	{\mcitemaxwidthsubitemform\space}
	{\relax}
	{\relax}
	
	\bibitem[Huang \latin{et~al.}(2017)Huang, Yuan, Shao, and Yan]{efficiency}
	Huang,~J.; Yuan,~Y.; Shao,~Y.; Yan,~Y. Understanding the physical properties of
	hybrid perovskites for photovoltaic applications. \emph{Nat. Rev. Mater.}
	\textbf{2017}, \emph{2}, 17042\relax
	\mciteBstWouldAddEndPuncttrue
	\mciteSetBstMidEndSepPunct{\mcitedefaultmidpunct}
	{\mcitedefaultendpunct}{\mcitedefaultseppunct}\relax
	\EndOfBibitem
	\bibitem[Deng \latin{et~al.}(2015)Deng, Wang, Yuan, and Huang]{doctorblade}
	Deng,~Y.; Wang,~Q.; Yuan,~Y.; Huang,~J. Vividly colorful hybrid perovskite
	solar cells by doctor-blade coating with perovskite photonic nanostructures.
	\emph{Mater. Horiz.} \textbf{2015}, \emph{2}, 578--583\relax
	\mciteBstWouldAddEndPuncttrue
	\mciteSetBstMidEndSepPunct{\mcitedefaultmidpunct}
	{\mcitedefaultendpunct}{\mcitedefaultseppunct}\relax
	\EndOfBibitem
	\bibitem[Steirer \latin{et~al.}(2016)Steirer, Schulz, Teeter, Stevanovic, Yang,
	Zhu, and Berry]{defect_insensitivity}
	Steirer,~K.~X.; Schulz,~P.; Teeter,~G.; Stevanovic,~V.; Yang,~M.; Zhu,~K.;
	Berry,~J.~J. Defect Tolerance in Methylammonium Lead Triiodide Perovskite.
	\emph{ACS Energy Lett.} \textbf{2016}, \emph{1}, 360--366\relax
	\mciteBstWouldAddEndPuncttrue
	\mciteSetBstMidEndSepPunct{\mcitedefaultmidpunct}
	{\mcitedefaultendpunct}{\mcitedefaultseppunct}\relax
	\EndOfBibitem
	\bibitem[Mantulnikovs \latin{et~al.}(2018)Mantulnikovs, Glushkova, Matus,
	Ćirić, Kollár, Forró, Horváth, and Sienkiewicz]{gazdetektor}
	Mantulnikovs,~K.; Glushkova,~A.; Matus,~P.; Ćirić,~L.; Kollár,~M.;
	Forró,~L.; Horváth,~E.; Sienkiewicz,~A. Morphology and Photoluminescence of
	$\text{CH}_3\text{NH}_3\text{PbI}_3$ Deposits on Nonplanar, Strongly Curved
	Substrates. \emph{ACS Photonics} \textbf{2018}, \emph{5}, 1476--1485\relax
	\mciteBstWouldAddEndPuncttrue
	\mciteSetBstMidEndSepPunct{\mcitedefaultmidpunct}
	{\mcitedefaultendpunct}{\mcitedefaultseppunct}\relax
	\EndOfBibitem
	\bibitem[Tan \latin{et~al.}(2014)Tan, Moghaddam, Lai, Docampo, Higler,
	Deschler, Price, Sadhanala, Pazos, Credgington, Hanusch, Bein, Snaith, and
	Friend]{dioda}
	Tan,~Z.-K.; Moghaddam,~R.~S.; Lai,~M.~L.; Docampo,~P.; Higler,~R.;
	Deschler,~F.; Price,~M.; Sadhanala,~A.; Pazos,~L.~M.; Credgington,~D.;
	Hanusch,~F.; Bein,~T.; Snaith,~H.~J.; Friend,~R.~H. Bright light-emitting
	diodes based on organometal halide perovskite. \emph{Nat. Nanotechnol.}
	\textbf{2014}, \emph{9}, 687--692\relax
	\mciteBstWouldAddEndPuncttrue
	\mciteSetBstMidEndSepPunct{\mcitedefaultmidpunct}
	{\mcitedefaultendpunct}{\mcitedefaultseppunct}\relax
	\EndOfBibitem
	\bibitem[Náfrádi \latin{et~al.}(2015)Náfrádi, Náfrádi, Forró, and
	Horváth]{rontgen}
	Náfrádi,~B.; Náfrádi,~G.; Forró,~L.; Horváth,~E. Methylammonium Lead
	Iodide for Efficient \text{X}-ray Energy Conversion. \emph{J. Phys. Chem. C}
	\textbf{2015}, \emph{119}, 25204--25208\relax
	\mciteBstWouldAddEndPuncttrue
	\mciteSetBstMidEndSepPunct{\mcitedefaultmidpunct}
	{\mcitedefaultendpunct}{\mcitedefaultseppunct}\relax
	\EndOfBibitem
	\bibitem[Ho-Baillie \latin{et~al.}(2022)Ho-Baillie, Sullivan, Bannerman,
	Talathi, Bing, Tang, Xu, Bhattacharyya, Cairns, and McKenzie]{spaceLAMI}
	Ho-Baillie,~A. W.~Y.; Sullivan,~H. G.~J.; Bannerman,~T.~A.; Talathi,~H.~P.;
	Bing,~J.; Tang,~S.; Xu,~A.; Bhattacharyya,~D.; Cairns,~I.~H.; McKenzie,~D.~R.
	Deployment Opportunities for Space Photovoltaics and the Prospects for
	Perovskite Solar Cells. \emph{Adv. Mater. Technol.} \textbf{2022}, \emph{7},
	2101059\relax
	\mciteBstWouldAddEndPuncttrue
	\mciteSetBstMidEndSepPunct{\mcitedefaultmidpunct}
	{\mcitedefaultendpunct}{\mcitedefaultseppunct}\relax
	\EndOfBibitem
	\bibitem[Green \latin{et~al.}(2022)Green, Dunlop, Hohl-Ebinger, Yoshita,
	Kopidakis, and Hao]{efficiency-uptodate}
	Green,~M.~A.; Dunlop,~E.~D.; Hohl-Ebinger,~J.; Yoshita,~M.; Kopidakis,~N.;
	Hao,~X. Solar cell efficiency tables (version 59). \emph{Progress in
		Photovoltaics: Research and Applications} \textbf{2022}, \emph{30},
	3--12\relax
	\mciteBstWouldAddEndPuncttrue
	\mciteSetBstMidEndSepPunct{\mcitedefaultmidpunct}
	{\mcitedefaultendpunct}{\mcitedefaultseppunct}\relax
	\EndOfBibitem
	\bibitem[Park and Zhu(2020)Park, and Zhu]{perovskitnapelem}
	Park,~N.-G.; Zhu,~K. Scalable fabrication and coating methods for perovskite
	solar cells and solar modules. \emph{Nat. Rev. Mater.} \textbf{2020},
	\emph{5}, 333--350\relax
	\mciteBstWouldAddEndPuncttrue
	\mciteSetBstMidEndSepPunct{\mcitedefaultmidpunct}
	{\mcitedefaultendpunct}{\mcitedefaultseppunct}\relax
	\EndOfBibitem
	\bibitem[Bhandari and Ellingson(2018)Bhandari, and
	Ellingson]{perovskitnapelem_ertekezes}
	Bhandari,~K.~P.; Ellingson,~R.~J. In \emph{A Comprehensive Guide to Solar
		Energy Systems}; Letcher,~T.~M., Fthenakis,~V.~M., Eds.; Academic Press,
	2018; pp 233 -- 254\relax
	\mciteBstWouldAddEndPuncttrue
	\mciteSetBstMidEndSepPunct{\mcitedefaultmidpunct}
	{\mcitedefaultendpunct}{\mcitedefaultseppunct}\relax
	\EndOfBibitem
	\bibitem[Fu \latin{et~al.}(2016)Fu, Feurer, Weiss, Pisoni, Avancini, Andres,
	Buecheler, and Tiwari]{perovskitnapelem_nature}
	Fu,~F.; Feurer,~T.; Weiss,~T.; Pisoni,~S.; Avancini,~E.; Andres,~C.;
	Buecheler,~S.; Tiwari,~A. High-efficiency inverted semi-transparent planar
	perovskite solar cells in substrate configuration. \emph{Nat. Energy}
	\textbf{2016}, \emph{2}, 16190\relax
	\mciteBstWouldAddEndPuncttrue
	\mciteSetBstMidEndSepPunct{\mcitedefaultmidpunct}
	{\mcitedefaultendpunct}{\mcitedefaultseppunct}\relax
	\EndOfBibitem
	\bibitem[Divitini \latin{et~al.}(2016)Divitini, Cacovich, Matteocci, Cin{\`a},
	Di~Carlo, and Ducati]{nature-temp-degradation}
	Divitini,~G.; Cacovich,~S.; Matteocci,~F.; Cin{\`a},~L.; Di~Carlo,~A.;
	Ducati,~C. In situ observation of heat-induced degradation of perovskite
	solar cells. \emph{Nat. Energy} \textbf{2016}, \emph{1}, 15012\relax
	\mciteBstWouldAddEndPuncttrue
	\mciteSetBstMidEndSepPunct{\mcitedefaultmidpunct}
	{\mcitedefaultendpunct}{\mcitedefaultseppunct}\relax
	\EndOfBibitem
	\bibitem[Arakcheeva \latin{et~al.}(2016)Arakcheeva, Chernyshov, Spina,
	Forr{\'{o}}, and
	Horv{\'{a}}th]{CH3NH3PbI3:precisestructuralconsequencesofwaterabsorptionatambientconditions}
	Arakcheeva,~A.; Chernyshov,~D.; Spina,~M.; Forr{\'{o}},~L.; Horv{\'{a}}th,~E.
	{CH${\sb 3}$NH${\sb 3}$PbI${\sb 3}$: precise structural consequences of water
		absorption at ambient conditions}. \emph{Acta Crystallogr., Sect. B}
	\textbf{2016}, \emph{72}, 716--722\relax
	\mciteBstWouldAddEndPuncttrue
	\mciteSetBstMidEndSepPunct{\mcitedefaultmidpunct}
	{\mcitedefaultendpunct}{\mcitedefaultseppunct}\relax
	\EndOfBibitem
	\bibitem[Aristidou \latin{et~al.}(2015)Aristidou, Sanchez-Molina,
	Chotchuangchutchaval, Brown, Martinez, Rath, and Haque]{oxigen-hatasa-2}
	Aristidou,~N.; Sanchez-Molina,~I.; Chotchuangchutchaval,~T.; Brown,~M.;
	Martinez,~L.; Rath,~T.; Haque,~S.~A. The Role of Oxygen in the Degradation of
	Methylammonium Lead Trihalide Perovskite Photoactive Layers. \emph{Angew.
		Chem., Int. Ed.} \textbf{2015}, \emph{54}, 8208--8212\relax
	\mciteBstWouldAddEndPuncttrue
	\mciteSetBstMidEndSepPunct{\mcitedefaultmidpunct}
	{\mcitedefaultendpunct}{\mcitedefaultseppunct}\relax
	\EndOfBibitem
	\bibitem[Poglitsch and Weber(1987)Poglitsch, and
	Weber]{dynamic-disorder-in-methylammoniumtrihalogenoplumbates-II-observed-by-milimeter-wave-spectroscopy}
	Poglitsch,~A.; Weber,~D. Dynamic disorder in methylammoniumtrihalogenoplumbates
	(II) observed by millimeter‐wave spectroscopy. \emph{J. Chem. Phys.}
	\textbf{1987}, \emph{87}, 6373--6378\relax
	\mciteBstWouldAddEndPuncttrue
	\mciteSetBstMidEndSepPunct{\mcitedefaultmidpunct}
	{\mcitedefaultendpunct}{\mcitedefaultseppunct}\relax
	\EndOfBibitem
	\bibitem[Ava \latin{et~al.}(2019)Ava, Al~Mamun, Marsillac, and
	Namkoong]{review_phasetransitions}
	Ava,~T.~T.; Al~Mamun,~A.; Marsillac,~S.; Namkoong,~G. A Review: Thermal
	Stability of Methylammonium Lead Halide Based Perovskite Solar Cells.
	\emph{Appl. Sci.} \textbf{2019}, \emph{9}, 188\relax
	\mciteBstWouldAddEndPuncttrue
	\mciteSetBstMidEndSepPunct{\mcitedefaultmidpunct}
	{\mcitedefaultendpunct}{\mcitedefaultseppunct}\relax
	\EndOfBibitem
	\bibitem[Chen \latin{et~al.}(2015)Chen, Foley, Ipek, Tyagi, Copley, Brown,
	Choi, and Lee]{allignment1}
	Chen,~T.; Foley,~B.~J.; Ipek,~B.; Tyagi,~M.; Copley,~J. R.~D.; Brown,~C.~M.;
	Choi,~J.~J.; Lee,~S.-H. Rotational dynamics of organic cations in the
	$\text{CH}_3\text{NH}_3\text{PbI}_3$ perovskite. \emph{Phys. Chem. Chem.
		Phys.} \textbf{2015}, \emph{17}, 31278--31286\relax
	\mciteBstWouldAddEndPuncttrue
	\mciteSetBstMidEndSepPunct{\mcitedefaultmidpunct}
	{\mcitedefaultendpunct}{\mcitedefaultseppunct}\relax
	\EndOfBibitem
	\bibitem[Bernard \latin{et~al.}(2018)Bernard, Wasylishen, Ratcliffe, Terskikh,
	Wu, Buriak, and Hauger]{allignemnt2}
	Bernard,~G.~M.; Wasylishen,~R.~E.; Ratcliffe,~C.~I.; Terskikh,~V.; Wu,~Q.;
	Buriak,~J.~M.; Hauger,~T. Methylammonium Cation Dynamics in Methylammonium
	Lead Halide Perovskites: A Solid-State NMR Perspective. \emph{J. Phys. Chem.
		A} \textbf{2018}, \emph{122}, 1560--1573\relax
	\mciteBstWouldAddEndPuncttrue
	\mciteSetBstMidEndSepPunct{\mcitedefaultmidpunct}
	{\mcitedefaultendpunct}{\mcitedefaultseppunct}\relax
	\EndOfBibitem
	\bibitem[Brivio \latin{et~al.}(2015)Brivio, Frost, Skelton, Jackson, Weber,
	Weller, Go\~ni, Leguy, Barnes, and Walsh]{allignment3}
	Brivio,~F.; Frost,~J.~M.; Skelton,~J.~M.; Jackson,~A.~J.; Weber,~O.~J.;
	Weller,~M.~T.; Go\~ni,~A.~R.; Leguy,~A. M.~A.; Barnes,~P. R.~F.; Walsh,~A.
	Lattice dynamics and vibrational spectra of the orthorhombic, tetragonal, and
	cubic phases of methylammonium lead iodide. \emph{Phys. Rev. B}
	\textbf{2015}, \emph{92}, 144308\relax
	\mciteBstWouldAddEndPuncttrue
	\mciteSetBstMidEndSepPunct{\mcitedefaultmidpunct}
	{\mcitedefaultendpunct}{\mcitedefaultseppunct}\relax
	\EndOfBibitem
	\bibitem[Weller \latin{et~al.}(2015)Weller, Weber, Henry, Di~Pumpo, and
	Hansen]{allignment4}
	Weller,~M.~T.; Weber,~O.~J.; Henry,~P.~F.; Di~Pumpo,~A.~M.; Hansen,~T.~C.
	Complete structure and cation orientation in the perovskite photovoltaic
	methylammonium lead iodide between 100 and 352 K. \emph{Chem. Commun.}
	\textbf{2015}, \emph{51}, 4180--4183\relax
	\mciteBstWouldAddEndPuncttrue
	\mciteSetBstMidEndSepPunct{\mcitedefaultmidpunct}
	{\mcitedefaultendpunct}{\mcitedefaultseppunct}\relax
	\EndOfBibitem
	\bibitem[Quarti \latin{et~al.}(2014)Quarti, Mosconi, and
	De~Angelis]{ferro-ion1}
	Quarti,~C.; Mosconi,~E.; De~Angelis,~F. Interplay of Orientational Order and
	Electronic Structure in Methylammonium Lead Iodide: Implications for Solar
	Cell Operation. \emph{Chem. Mater.} \textbf{2014}, \emph{26},
	6557--6569\relax
	\mciteBstWouldAddEndPuncttrue
	\mciteSetBstMidEndSepPunct{\mcitedefaultmidpunct}
	{\mcitedefaultendpunct}{\mcitedefaultseppunct}\relax
	\EndOfBibitem
	\bibitem[Frost \latin{et~al.}(2014)Frost, Butler, Brivio, Hendon, van
	Schilfgaarde, and Walsh]{ferro-ion2}
	Frost,~J.~M.; Butler,~K.~T.; Brivio,~F.; Hendon,~C.~H.; van Schilfgaarde,~M.;
	Walsh,~A. Atomistic Origins of High-Performance in Hybrid Halide Perovskite
	Solar Cells. \emph{Nano Lett.} \textbf{2014}, \emph{14}, 2584--2590\relax
	\mciteBstWouldAddEndPuncttrue
	\mciteSetBstMidEndSepPunct{\mcitedefaultmidpunct}
	{\mcitedefaultendpunct}{\mcitedefaultseppunct}\relax
	\EndOfBibitem
	\bibitem[Kim \latin{et~al.}(2015)Kim, Kim, Kim, Shin, Gupta, Jung, Kim, and
	Park]{ferro-ion3}
	Kim,~H.-S.; Kim,~S.~K.; Kim,~B.~J.; Shin,~K.-S.; Gupta,~M.~K.; Jung,~H.~S.;
	Kim,~S.-W.; Park,~N.-G. Ferroelectric Polarization in CH$_3$NH$_3$PbI$_3$
	Perovskite. \emph{J. Phys. Chem. Lett.} \textbf{2015}, \emph{6},
	1729--1735\relax
	\mciteBstWouldAddEndPuncttrue
	\mciteSetBstMidEndSepPunct{\mcitedefaultmidpunct}
	{\mcitedefaultendpunct}{\mcitedefaultseppunct}\relax
	\EndOfBibitem
	\bibitem[Kutes \latin{et~al.}(2014)Kutes, Ye, Zhou, Pang, Huey, and
	Padture]{ferro-ion4}
	Kutes,~Y.; Ye,~L.; Zhou,~Y.; Pang,~S.; Huey,~B.~D.; Padture,~N.~P. Direct
	Observation of Ferroelectric Domains in Solution-Processed
	CH$_3$NH$_3$PbI$_3$ Perovskite Thin Films. \emph{J. Phys. Chem. Lett.}
	\textbf{2014}, \emph{5}, 3335--3339\relax
	\mciteBstWouldAddEndPuncttrue
	\mciteSetBstMidEndSepPunct{\mcitedefaultmidpunct}
	{\mcitedefaultendpunct}{\mcitedefaultseppunct}\relax
	\EndOfBibitem
	\bibitem[Beilsten-Edmands \latin{et~al.}(2015)Beilsten-Edmands, Eperon,
	Johnson, Snaith, and Radaelli]{ferro-ion5}
	Beilsten-Edmands,~J.; Eperon,~G.~E.; Johnson,~R.~D.; Snaith,~H.~J.;
	Radaelli,~P.~G. Non-ferroelectric nature of the conductance hysteresis in
	CH$_3$NH$_3$PbI$_3$ perovskite-based photovoltaic devices. \emph{Appl. Phys.
		Lett.} \textbf{2015}, \emph{106}, 173502\relax
	\mciteBstWouldAddEndPuncttrue
	\mciteSetBstMidEndSepPunct{\mcitedefaultmidpunct}
	{\mcitedefaultendpunct}{\mcitedefaultseppunct}\relax
	\EndOfBibitem
	\bibitem[{Kunst} and {Beck}(1986){Kunst}, and {Beck}]{kunst1}
	{Kunst},~M.; {Beck},~G. {The study of charge carrier kinetics in semiconductors
		by microwave conductivity measurements}. \emph{J. Appl. Phys.} \textbf{1986},
	\emph{60}, 3558--3566\relax
	\mciteBstWouldAddEndPuncttrue
	\mciteSetBstMidEndSepPunct{\mcitedefaultmidpunct}
	{\mcitedefaultendpunct}{\mcitedefaultseppunct}\relax
	\EndOfBibitem
	\bibitem[Kunst and Beck(1988)Kunst, and Beck]{kunst2}
	Kunst,~M.; Beck,~G. The study of charge carrier kinetics in semiconductors by
	microwave conductivity measurements. $\text{II}$. \emph{J. Appl. Phys.}
	\textbf{1988}, \emph{63}, 1093--1098\relax
	\mciteBstWouldAddEndPuncttrue
	\mciteSetBstMidEndSepPunct{\mcitedefaultmidpunct}
	{\mcitedefaultendpunct}{\mcitedefaultseppunct}\relax
	\EndOfBibitem
	\bibitem[{Sinton} \latin{et~al.}(1996){Sinton}, {Cuevas}, and
	{Stuckings}]{Sinton1}
	{Sinton},~R.~A.; {Cuevas},~A.; {Stuckings},~M. Quasi-steady-state
	photoconductance, a new method for solar cell material and device
	characterization. Conference Record of the Twenty Fifth IEEE Photovoltaic
	Specialists Conference - 1996. 1996; pp 457--460\relax
	\mciteBstWouldAddEndPuncttrue
	\mciteSetBstMidEndSepPunct{\mcitedefaultmidpunct}
	{\mcitedefaultendpunct}{\mcitedefaultseppunct}\relax
	\EndOfBibitem
	\bibitem[Goodarzi \latin{et~al.}(2019)Goodarzi, Sinton, and Macdonald]{sinton2}
	Goodarzi,~M.; Sinton,~R.; Macdonald,~D. Quasi-steady-state photoconductance
	bulk lifetime measurements on silicon ingots with deeper photogeneration.
	\emph{AIP Adv.} \textbf{2019}, \emph{9}, 015128\relax
	\mciteBstWouldAddEndPuncttrue
	\mciteSetBstMidEndSepPunct{\mcitedefaultmidpunct}
	{\mcitedefaultendpunct}{\mcitedefaultseppunct}\relax
	\EndOfBibitem
	\bibitem[Gyüre-Garami \latin{et~al.}(2019)Gyüre-Garami, Blum, Sági, Bojtor,
	Kollarics, Csősz, Márkus, Volk, and Simon]{gyregarami2019ultrafast}
	Gyüre-Garami,~B.; Blum,~B.; Sági,~O.; Bojtor,~A.; Kollarics,~S.; Csősz,~G.;
	Márkus,~B.~G.; Volk,~J.; Simon,~F. Ultrafast sensing of photoconductivity
	decay using microwave resonators. \emph{J. Appl. Phys.} \textbf{2019},
	\emph{126}, 235702\relax
	\mciteBstWouldAddEndPuncttrue
	\mciteSetBstMidEndSepPunct{\mcitedefaultmidpunct}
	{\mcitedefaultendpunct}{\mcitedefaultseppunct}\relax
	\EndOfBibitem
	\bibitem[Sólyom(2008)]{solyom2}
	Sólyom,~J. \emph{Fundamentals of the physics of solids II: Electronic
		Properties}; Springer, 2008\relax
	\mciteBstWouldAddEndPuncttrue
	\mciteSetBstMidEndSepPunct{\mcitedefaultmidpunct}
	{\mcitedefaultendpunct}{\mcitedefaultseppunct}\relax
	\EndOfBibitem
	\bibitem[Osherov \latin{et~al.}(2016)Osherov, Hutter, Galkowski, Brenes, Maude,
	Nicholas, Plochocka, Bulović, Savenije, and Stranks]{structural_hyst}
	Osherov,~A.; Hutter,~E.~M.; Galkowski,~K.; Brenes,~R.; Maude,~D.~K.;
	Nicholas,~R.~J.; Plochocka,~P.; Bulović,~V.; Savenije,~T.~J.; Stranks,~S.~D.
	The Impact of Phase Retention on the Structural and Optoelectronic Properties
	of Metal Halide Perovskites. \emph{Adv. Mater.} \textbf{2016}, \emph{28},
	10757--10763\relax
	\mciteBstWouldAddEndPuncttrue
	\mciteSetBstMidEndSepPunct{\mcitedefaultmidpunct}
	{\mcitedefaultendpunct}{\mcitedefaultseppunct}\relax
	\EndOfBibitem
	\bibitem[Gélvez-Rueda \latin{et~al.}(2016)Gélvez-Rueda, Cao, Patwardhan,
	Renaud, Stoumpos, Schatz, Hupp, Farha, Savenije, Kanatzidis, and
	Grozema]{built-in-E}
	Gélvez-Rueda,~M.~C.; Cao,~D.~H.; Patwardhan,~S.; Renaud,~N.; Stoumpos,~C.~C.;
	Schatz,~G.~C.; Hupp,~J.~T.; Farha,~O.~K.; Savenije,~T.~J.; Kanatzidis,~M.~G.;
	Grozema,~F.~C. Effect of Cation Rotation on Charge Dynamics in Hybrid Lead
	Halide Perovskites. \emph{J. Phys. Chem. C} \textbf{2016}, \emph{120},
	16577--16585\relax
	\mciteBstWouldAddEndPuncttrue
	\mciteSetBstMidEndSepPunct{\mcitedefaultmidpunct}
	{\mcitedefaultendpunct}{\mcitedefaultseppunct}\relax
	\EndOfBibitem
	\bibitem[Leguy \latin{et~al.}({2015})Leguy, Frost, McMahon, Sakai, Kockelmann,
	Law, Li, Foglia, Walsh, O'Regan, Nelson, Cabral, and
	Barnes]{Leguy_MA_ordering}
	Leguy,~A. M.~A.; Frost,~J.~M.; McMahon,~A.~P.; Sakai,~V.~G.; Kockelmann,~W.;
	Law,~C.; Li,~X.; Foglia,~F.; Walsh,~A.; O'Regan,~B.~C.; Nelson,~J.;
	Cabral,~J.~T.; Barnes,~P. R.~F. {The dynamics of methylammonium ions in
		hybrid organic-inorganic perovskite solar cells}. \emph{{Nat. Commun.}}
	\textbf{{2015}}, \emph{{6}}, {7124}\relax
	\mciteBstWouldAddEndPuncttrue
	\mciteSetBstMidEndSepPunct{\mcitedefaultmidpunct}
	{\mcitedefaultendpunct}{\mcitedefaultseppunct}\relax
	\EndOfBibitem
	\bibitem[Semonin \latin{et~al.}(2016)Semonin, Elbaz, Straus, Hull, Paley,
	van~der Zande, Hone, Kymissis, Kagan, Roy, and Owen]{TRMCD_10us}
	Semonin,~O.~E.; Elbaz,~G.~A.; Straus,~D.~B.; Hull,~T.~D.; Paley,~D.~W.; van~der
	Zande,~A.~M.; Hone,~J.~C.; Kymissis,~I.; Kagan,~C.~R.; Roy,~X.; Owen,~J.~S.
	Limits of Carrier Diffusion in n-Type and p-Type CH$_3$NH$_3$PbI$_3$
	Perovskite Single Crystals. \emph{J. Phys. Chem. Lett.} \textbf{2016},
	\emph{7}, 3510--3518\relax
	\mciteBstWouldAddEndPuncttrue
	\mciteSetBstMidEndSepPunct{\mcitedefaultmidpunct}
	{\mcitedefaultendpunct}{\mcitedefaultseppunct}\relax
	\EndOfBibitem
	\bibitem[Labram \latin{et~al.}(2018)Labram, Perry, Venkatesan, and
	Chabinyc]{TRMC_SS}
	Labram,~J.~G.; Perry,~E.~E.; Venkatesan,~N.~R.; Chabinyc,~M.~L. Steady-state
	microwave conductivity reveals mobility-lifetime product in methylammonium
	lead iodide. \emph{Appl. Phys. Lett.} \textbf{2018}, \emph{113}, 153902\relax
	\mciteBstWouldAddEndPuncttrue
	\mciteSetBstMidEndSepPunct{\mcitedefaultmidpunct}
	{\mcitedefaultendpunct}{\mcitedefaultseppunct}\relax
	\EndOfBibitem
	\bibitem[Hutter \latin{et~al.}(2015)Hutter, Eperon, Stranks, and
	Savenije]{TRMC_cavity}
	Hutter,~E.~M.; Eperon,~G.~E.; Stranks,~S.~D.; Savenije,~T.~J. Charge Carriers
	in Planar and Meso-Structured Organic–Inorganic Perovskites: Mobilities,
	Lifetimes, and Concentrations of Trap States. \emph{J. Phys. Chem. Lett.}
	\textbf{2015}, \emph{6}, 3082--3090\relax
	\mciteBstWouldAddEndPuncttrue
	\mciteSetBstMidEndSepPunct{\mcitedefaultmidpunct}
	{\mcitedefaultendpunct}{\mcitedefaultseppunct}\relax
	\EndOfBibitem
	\bibitem[Chattopadhyay \latin{et~al.}(2020)Chattopadhyay, Kokenyesi, Hong,
	Watts, and Labram]{TRMC_cavity2}
	Chattopadhyay,~S.; Kokenyesi,~R.~S.; Hong,~M.~J.; Watts,~C.~L.; Labram,~J.~G.
	Resolving in-plane and out-of-plane mobility using time resolved microwave
	conductivity. \emph{J. Mater. Chem. C} \textbf{2020}, \emph{8},
	10761--10766\relax
	\mciteBstWouldAddEndPuncttrue
	\mciteSetBstMidEndSepPunct{\mcitedefaultmidpunct}
	{\mcitedefaultendpunct}{\mcitedefaultseppunct}\relax
	\EndOfBibitem
	\bibitem[Hutter \latin{et~al.}(2017)Hutter, G{\'e}lvez-Rueda, Osherov,
	Bulovi{\'{c}}, Grozema, Stranks, and Savenije]{TRMC_thin_film}
	Hutter,~E.~M.; G{\'e}lvez-Rueda,~M.~C.; Osherov,~A.; Bulovi{\'{c}},~V.;
	Grozema,~F.~C.; Stranks,~S.~D.; Savenije,~T.~J. Direct--indirect character of
	the bandgap in methylammonium lead iodide perovskite. \emph{Nat. Mater.}
	\textbf{2017}, \emph{16}, 115--120\relax
	\mciteBstWouldAddEndPuncttrue
	\mciteSetBstMidEndSepPunct{\mcitedefaultmidpunct}
	{\mcitedefaultendpunct}{\mcitedefaultseppunct}\relax
	\EndOfBibitem
	\bibitem[Hong \latin{et~al.}(2019)Hong, Svadlenak, Goulas, and
	Labram]{TRMC_thin_film2}
	Hong,~M.~J.; Svadlenak,~S.~R.; Goulas,~K.~A.; Labram,~J.~G. Thermal stability
	of mobility in methylammonium lead iodide. \emph{J. Phys Mater.}
	\textbf{2019}, \emph{3}, 014003\relax
	\mciteBstWouldAddEndPuncttrue
	\mciteSetBstMidEndSepPunct{\mcitedefaultmidpunct}
	{\mcitedefaultendpunct}{\mcitedefaultseppunct}\relax
	\EndOfBibitem
	\bibitem[Bi \latin{et~al.}(2016)Bi, Hutter, Fang, Dong, Huang, and
	Savenije]{TRMC_transmission}
	Bi,~Y.; Hutter,~E.~M.; Fang,~Y.; Dong,~Q.; Huang,~J.; Savenije,~T.~J. Charge
	Carrier Lifetimes Exceeding $15\ \mu\text{s}$ in Methylammonium Lead Iodide
	Single Crystals. \emph{J. Phys. Chem. Lett.} \textbf{2016}, \emph{7},
	923--928\relax
	\mciteBstWouldAddEndPuncttrue
	\mciteSetBstMidEndSepPunct{\mcitedefaultmidpunct}
	{\mcitedefaultendpunct}{\mcitedefaultseppunct}\relax
	\EndOfBibitem
	\bibitem[Li \latin{et~al.}(2016)Li, Wang, Cheng, Chen, Wu, Liu, Huang, and
	Duan]{temp-mobility}
	Li,~D.; Wang,~G.; Cheng,~H.-C.; Chen,~C.-Y.; Wu,~H.; Liu,~Y.; Huang,~Y.;
	Duan,~X. Size-dependent phase transition in methylammonium lead iodide
	perovskite microplate crystals. \emph{Nat. Commun.} \textbf{2016}, \emph{7},
	11330\relax
	\mciteBstWouldAddEndPuncttrue
	\mciteSetBstMidEndSepPunct{\mcitedefaultmidpunct}
	{\mcitedefaultendpunct}{\mcitedefaultseppunct}\relax
	\EndOfBibitem
	\bibitem[Wehrenfennig \latin{et~al.}(2014)Wehrenfennig, Liu, Snaith, Johnston,
	and Herz]{temp-PL}
	Wehrenfennig,~C.; Liu,~M.; Snaith,~H.~J.; Johnston,~M.~B.; Herz,~L.~M. Charge
	carrier recombination channels in the low-temperature phase of
	organic-inorganic lead halide perovskite thin films. \emph{APL Mater.}
	\textbf{2014}, \emph{2}, 081513\relax
	\mciteBstWouldAddEndPuncttrue
	\mciteSetBstMidEndSepPunct{\mcitedefaultmidpunct}
	{\mcitedefaultendpunct}{\mcitedefaultseppunct}\relax
	\EndOfBibitem
	\bibitem[Lappas \latin{et~al.}(1995)Lappas, Kosaka, Tanigaki, and
	Prassides]{lappas_quench}
	Lappas,~A.; Kosaka,~M.; Tanigaki,~K.; Prassides,~K. An Orientationally-Ordered
	Primitive-Cubic Form of the Fulleride CsC$_{60}$. \emph{J. Am. Chem. Soc.}
	\textbf{1995}, \emph{117}, 7560--7561\relax
	\mciteBstWouldAddEndPuncttrue
	\mciteSetBstMidEndSepPunct{\mcitedefaultmidpunct}
	{\mcitedefaultendpunct}{\mcitedefaultseppunct}\relax
	\EndOfBibitem
	\bibitem[Janes \latin{et~al.}(1998)Janes, Edge, Robinson, Allen, and
	Thompson]{Janes1998}
	Janes,~R.; Edge,~M.; Robinson,~J.; Allen,~N.~S.; Thompson,~F. Microwave
	photodielectric and photoconductivity studies of commercial titanium dioxide
	pigments: the influence of transition metal dopants. \emph{J. Mater. Sci.}
	\textbf{1998}, \emph{33}, 3031--3036\relax
	\mciteBstWouldAddEndPuncttrue
	\mciteSetBstMidEndSepPunct{\mcitedefaultmidpunct}
	{\mcitedefaultendpunct}{\mcitedefaultseppunct}\relax
	\EndOfBibitem
	\bibitem[Savenije \latin{et~al.}(2013)Savenije, Ferguson, Kopidakis, and
	Rumbles]{Savenije-cavitysetup}
	Savenije,~T.~J.; Ferguson,~A.~J.; Kopidakis,~N.; Rumbles,~G. Revealing the
	Dynamics of Charge Carriers in Polymer:Fullerene Blends Using Photoinduced
	Time-Resolved Microwave Conductivity. \emph{J. Phys. Chem. C} \textbf{2013},
	\emph{117}, 24085--24103\relax
	\mciteBstWouldAddEndPuncttrue
	\mciteSetBstMidEndSepPunct{\mcitedefaultmidpunct}
	{\mcitedefaultendpunct}{\mcitedefaultseppunct}\relax
	\EndOfBibitem
	\bibitem[Savenije \latin{et~al.}(2020)Savenije, Guo, Caselli, and
	Hutter]{savenije_MHPTRMC}
	Savenije,~T.~J.; Guo,~D.; Caselli,~V.~M.; Hutter,~E.~M. Quantifying
	Charge-Carrier Mobilities and Recombination Rates in Metal Halide Perovskites
	from Time-Resolved Microwave Photoconductivity Measurements. \emph{Adv.
		Energy Mater.} \textbf{2020}, \emph{10}, 1903788\relax
	\mciteBstWouldAddEndPuncttrue
	\mciteSetBstMidEndSepPunct{\mcitedefaultmidpunct}
	{\mcitedefaultendpunct}{\mcitedefaultseppunct}\relax
	\EndOfBibitem
	\bibitem[Csősz \latin{et~al.}(2018)Csősz, Márkus, Jánossy, Nemes, Murányi,
	Klupp, Kamarás, Kogan, Bud’ko, Canfield, and Simon]{csosz_cavity}
	Csősz,~G.; Márkus,~B.~G.; Jánossy,~A.; Nemes,~N.~M.; Murányi,~F.;
	Klupp,~G.; Kamarás,~K.; Kogan,~V.~G.; Bud’ko,~S.~L.; Canfield,~P.~C.;
	Simon,~F. Giant microwave absorption in fine powders of superconductors.
	\emph{Sci. Rep.} \textbf{2018}, \emph{8}, 11480--11488\relax
	\mciteBstWouldAddEndPuncttrue
	\mciteSetBstMidEndSepPunct{\mcitedefaultmidpunct}
	{\mcitedefaultendpunct}{\mcitedefaultseppunct}\relax
	\EndOfBibitem
	\bibitem[Novikov \latin{et~al.}(2010)Novikov, Marinin, and Rabenok]{cavity1}
	Novikov,~G.~F.; Marinin,~A.~A.; Rabenok,~E.~V. Microwave measurements of the
	pulsed photoconductivity and photoelectric effect. \emph{Instrum. Exp. Tech.}
	\textbf{2010}, \emph{53}, 233--239\relax
	\mciteBstWouldAddEndPuncttrue
	\mciteSetBstMidEndSepPunct{\mcitedefaultmidpunct}
	{\mcitedefaultendpunct}{\mcitedefaultseppunct}\relax
	\EndOfBibitem
	\bibitem[Joubert \latin{et~al.}(2004)Joubert, Kazanskii, Guyot, G\^acon, and
	P\'edrini]{cavity2}
	Joubert,~M.-F.; Kazanskii,~S.~A.; Guyot,~Y.; G\^acon,~J.-C.; P\'edrini,~C.
	Microwave study of photoconductivity induced by laser pulses in
	rare-earth-doped dielectric crystals. \emph{Phys. Rev. B} \textbf{2004},
	\emph{69}, 165217\relax
	\mciteBstWouldAddEndPuncttrue
	\mciteSetBstMidEndSepPunct{\mcitedefaultmidpunct}
	{\mcitedefaultendpunct}{\mcitedefaultseppunct}\relax
	\EndOfBibitem
	\bibitem[Mantulnikovs \latin{et~al.}(2020)Mantulnikovs, Szirmai, Koll{\'a}r,
	Stevens, Andri{\v{c}}evi{\'c}, Glushkova, Rossi, Bugnon, Horv{\'a}th,
	Sienkiewicz, \latin{et~al.} others]{mantulnikovs2020light}
	Mantulnikovs,~K.; Szirmai,~P.; Koll{\'a}r,~M.; Stevens,~J.;
	Andri{\v{c}}evi{\'c},~P.; Glushkova,~A.; Rossi,~L.; Bugnon,~P.;
	Horv{\'a}th,~E.; Sienkiewicz,~A., \latin{et~al.}  Light-induced charge
	transfer at the CH$_3$NH$_3$PbI$_3$/TiO$_2$ interface—a low-temperature
	photo-electron paramagnetic resonance assay. \emph{JPhys Photonics}
	\textbf{2020}, \emph{2}, 014007\relax
	\mciteBstWouldAddEndPuncttrue
	\mciteSetBstMidEndSepPunct{\mcitedefaultmidpunct}
	{\mcitedefaultendpunct}{\mcitedefaultseppunct}\relax
	\EndOfBibitem
	\bibitem[Mosconi \latin{et~al.}(2016)Mosconi, Umari, and
	De~Angelis]{absorptions_lami}
	Mosconi,~E.; Umari,~P.; De~Angelis,~F. Electronic and optical properties of
	MAPbX$_3$ perovskites (X= I, Br, Cl): a unified DFT and GW theoretical
	analysis. \emph{Phys. Chem. Chem. Phys.} \textbf{2016}, \emph{18},
	27158--27164\relax
	\mciteBstWouldAddEndPuncttrue
	\mciteSetBstMidEndSepPunct{\mcitedefaultmidpunct}
	{\mcitedefaultendpunct}{\mcitedefaultseppunct}\relax
	\EndOfBibitem
	\bibitem[Maculan \latin{et~al.}(2015)Maculan, Sheikh, Abdelhady, Saidaminov,
	Haque, Murali, Alarousu, Mohammed, Wu, and Bakr]{lami_absorption}
	Maculan,~G.; Sheikh,~A.~D.; Abdelhady,~A.~L.; Saidaminov,~M.~I.; Haque,~M.~A.;
	Murali,~B.; Alarousu,~E.; Mohammed,~O.~F.; Wu,~T.; Bakr,~O.~M.
	CH$_3$NH$_3$PbCl$_3$ single crystals: inverse temperature crystallization and
	visible-blind UV-photodetector. \emph{J. Phys. Chem. Lett.} \textbf{2015},
	\emph{6}, 3781--3786\relax
	\mciteBstWouldAddEndPuncttrue
	\mciteSetBstMidEndSepPunct{\mcitedefaultmidpunct}
	{\mcitedefaultendpunct}{\mcitedefaultseppunct}\relax
	\EndOfBibitem
	\bibitem[Pozar(2011)]{pozar}
	Pozar,~D. \emph{Microwave Engineering - Solutions Manual}; Wiley, 2011; Vol. 4
	ed.\relax
	\mciteBstWouldAddEndPunctfalse
	\mciteSetBstMidEndSepPunct{\mcitedefaultmidpunct}
	{}{\mcitedefaultseppunct}\relax
	\EndOfBibitem
	\bibitem[Chen \latin{et~al.}(2004)Chen, Ong, Neo, Varadan, and Varadan]{LFChen}
	Chen,~L.; Ong,~C.; Neo,~C.; Varadan,~V.; Varadan,~V. \emph{Microwave
		Electronics: Measurement and Materials Characterization}; Wiley, 2004\relax
	\mciteBstWouldAddEndPuncttrue
	\mciteSetBstMidEndSepPunct{\mcitedefaultmidpunct}
	{\mcitedefaultendpunct}{\mcitedefaultseppunct}\relax
	\EndOfBibitem
	\bibitem[Klein \latin{et~al.}(1993)Klein, Donovan, Dressel, and
	Gr{\"u}ner]{Klein}
	Klein,~O.; Donovan,~S.; Dressel,~M.; Gr{\"u}ner,~G. Microwave cavity
	perturbation technique: Part I: Principles. \emph{Int. J. Infrared Millimeter
		Waves} \textbf{1993}, \emph{14}, 2423--2457\relax
	\mciteBstWouldAddEndPuncttrue
	\mciteSetBstMidEndSepPunct{\mcitedefaultmidpunct}
	{\mcitedefaultendpunct}{\mcitedefaultseppunct}\relax
	\EndOfBibitem
	\bibitem[Landau \latin{et~al.}(2013)Landau, Bell, Kearsley, Pitaevskii,
	Lifshitz, and Sykes]{landau2013electrodynamics}
	Landau,~L.; Bell,~J.; Kearsley,~M.; Pitaevskii,~L.; Lifshitz,~E.; Sykes,~J.
	\emph{Electrodynamics of Continuous Media}; Course Of Theoretical Physics;
	Elsevier Science, 2013\relax
	\mciteBstWouldAddEndPuncttrue
	\mciteSetBstMidEndSepPunct{\mcitedefaultmidpunct}
	{\mcitedefaultendpunct}{\mcitedefaultseppunct}\relax
	\EndOfBibitem
	\bibitem[Poole(1983)]{PooleBook}
	Poole,~C.~P. \emph{Electron Spin Resonance}, 1st ed.; John Wiley \& Sons: New
	York, 1983\relax
	\mciteBstWouldAddEndPuncttrue
	\mciteSetBstMidEndSepPunct{\mcitedefaultmidpunct}
	{\mcitedefaultendpunct}{\mcitedefaultseppunct}\relax
	\EndOfBibitem
\end{mcitethebibliography}


\providecommand{\latin}[1]{#1}
\makeatletter
\providecommand{\doi}
{\begingroup\let\do\@makeother\dospecials
	\catcode`\{=1 \catcode`\}=2 \doi@aux}
\providecommand{\doi@aux}[1]{\endgroup\texttt{#1}}
\makeatother
\providecommand*\mcitethebibliography{\thebibliography}
\csname @ifundefined\endcsname{endmcitethebibliography}
{\let\endmcitethebibliography\endthebibliography}{}

\end{document}